\documentclass[a4paper, 11pt]{article}
\pdfoutput=1
\usepackage{mathrsfs,dsfont,threeparttable,amsthm,appendix,multirow}
\usepackage{jheppub}
\usepackage{amsmath}
\usepackage{amssymb}
\usepackage{graphicx,xcolor,subfigure}
\usepackage{empheq}
\usepackage{esint}
\usepackage{hyperref}
\usepackage[shortlabels]{enumitem}
\usepackage{tikz, pgf} 
\usetikzlibrary{arrows}
\usepackage{float}
\usepackage{slashed}
\usepackage{array}
\usepackage[normalem]{ulem}
\usepackage{tikz}
\usetikzlibrary{shapes.misc}
\tikzset{cross/.style={cross out, draw=black, minimum size=2*(#1-\pgflinewidth), inner sep=0pt, outer sep=0pt},
cross/.default={3pt}}
\usetikzlibrary{decorations.pathmorphing}
\tikzset{snake it/.style={decorate, decoration=snake}}
\usepackage{subfigure,tikz}
\newcommand{\be}{\begin{equation}}
\newcommand{\ee}{\end{equation}}

\newtheorem*{thm}{Theorem}
\title{Non-invertible twisted compactification of class $\mathcal S$ theory and $(B,B,B)$ branes}
\author{Yankun Ma}
\affiliation{School of Physics and Astronomy, Sun Yat-Sen University}
\emailAdd{mayk8@mail2.sysu.edu.cn}
\abstract{We study \emph{non-invertible twisted compactification} of class $\mathcal S$ theories on $S^1$: we insert a non-invertible symmetry defect at $S^1$ extending along remaining directions and then compactify on $S^1$. We show that the resulting 3d theory is 3d $\mathcal N=4$ sigma model whose target space is a hyperK\"ahler submanifold of Hitchin moduli space, i.e. a $(B,B,B)$ brane. The $(B,B,B)$ brane is the fixed point set on Hitchin moduli space of a finite subgroup of mapping class group of underlying Riemann surface.  We describe the $(B,B,B)$ branes as affine varieties and calculate concrete examples of these $(B,B,B)$ branes for type $A_1$, genus $2$ class $\mathcal S$ theory.}

\begin{document}
\maketitle
\section{Introduction}

Symmetry is a central theme in Physics. Recently there has been growing interests in studying \emph{generalized global symmetries} in quantum field theory. Central to the revolution is the identification of symmetry as topological defects \cite{Gaiotto:2014kfa}. In this language, the ordinary symmetry is a codimension-1 topological operator $U_g(\mathcal M^{d-1})$ whose fusion forms group laws. Exhibiting similar properties as ordinary global symmetry, generalized symmetries also possess structures deviating from standard ones. This leads to the notion of higher-form symmetry, higher-group symmetry and non-invertible symmetry.

	Non-invertible symmetry has been studied in $2d$ RCFT for many decades, known as topological defect lines \cite{Verlinde:1988sn,Frohlich:2006ch}. The fusion and junction of these TDLs are not described by a group but rather a fusion category \cite{Chang:2018iay}.  A prototypical example is the Kramers-Wannier line that implements Kramers-Wannier duality \cite{Frohlich:2004ef}. However it's a more recent result that non-invertible symmetry also exists in $d>2$ QFTs \cite{Kaidi:2021xfk,Choi:2021kmx,Roumpedakis:2022aik}. Among the many constructions, in this paper we will focus on the \emph{non-invertible self-duality defect} \cite{Choi:2021kmx}, which is a generalization of Kramers-Wannier line.

\begin{figure}[tbp]
\begin{center}
\[\hspace*{-0.8 cm}\begin{tikzpicture}[baseline=19,scale=0.8]
\draw[  thick] (0,-0.2)--(0,3.2);
\draw[  thick] (3,-0.2)--(3,3.2);
  
          \shade[line width=2pt, top color=red,opacity=0.4] 
    (0,0) to [out=90, in=-90]  (0,3)
    to [out=0,in=180] (3,3)
    to [out = -90, in =90] (3,0)
    to [out=190, in =0]  (0,0);
    
\node[left] at (-0.7,1.5) {$\mathcal T_\rho(x)$};
    \node at (1.5,1.5) {$\mathcal T_{\mathsf S \cdot \rho}(\mathsf S\cdot x)$};
    \node[right] at (3.7,1.5) {$\mathcal T_{\sigma\mathsf S\cdot\rho}(\mathsf S\cdot x)$};
      \node[left] at (0,-0) {$\mathsf S$};
          \node[right] at (3,-0) {$\sigma$};  
\end{tikzpicture}
\hspace{0.35 in}\Rightarrow\hspace{0.35 in}
\begin{tikzpicture}[baseline=19,scale=0.8]
\draw[red, thick] (0,-0.2)--(0,3.2);
  \node[left] at (-0.7,1.5) {$\mathcal T_\rho(x)$};
    \node[right] at (+0.7,1.5) {$\mathcal T_{\sigma\mathsf S\cdot\rho}(\mathsf S\cdot x)$};
      \node[left] at (0,0) {$\mathcal N$};
\end{tikzpicture}
\]
\caption{Combining the duality action $\mathsf S$ with a topological manipulation $\sigma$ yields a new topological interface $\mathcal N$, which can be made a non-invertible symmetry defect. }
\label{fig:Ndef}
\end{center}
\end{figure}
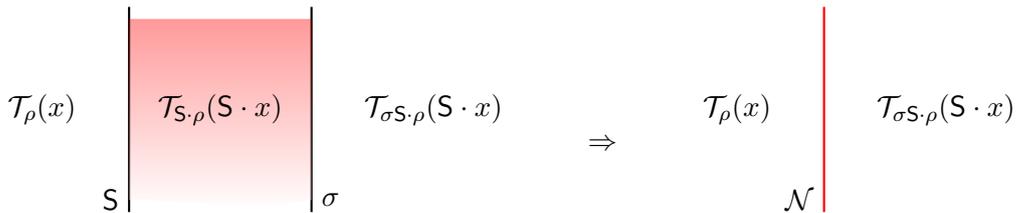

It's constructed as follows. Denote a family of theories by $\mathcal T_{\rho}(x)$, where $x\in \mathfrak M$ parametrizes the conformal manifold and $\rho$ labels a possible global structure. The duality group $\Gamma$, which acts on the conformal manifold $\mathfrak M$, also changes the global structure. Its action is implemented by a topological interface $\mathsf S:\mathcal T_\rho(x)\mapsto \mathcal T_{\mathsf S\cdot \rho}(\mathsf S\cdot x) \quad \mathsf S\in \Gamma$. To map the system back to itself there's also a topological manipulation which only changes the global form: $\sigma: \mathcal T_\rho(x)\mapsto \mathcal T_{\sigma\cdot \rho}(x)$. Generally it consists of gauging a $q$-form global symmetry and stacking an SPT phase. Combining the two topological interfaces together, we produce a new topological interface $\mathcal N:=\sigma\mathsf S$, as depicted in Fig. \ref{fig:Ndef}.

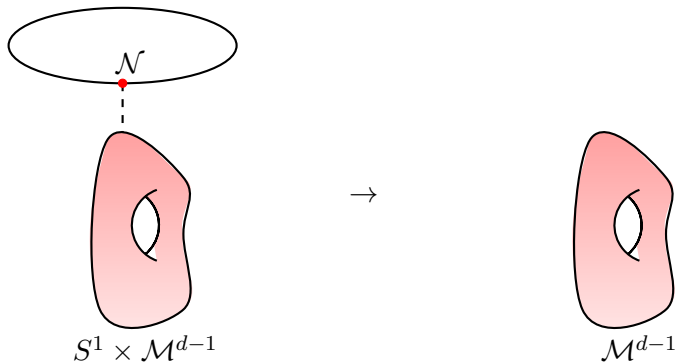
\begin{figure}[tbp]
\begin{center}

  \begin{tikzpicture}
\draw[thick] (0,0) circle [x radius=1.5cm, y radius=0.5cm];
\node[above] at (0.1,-0.5) {$\mathcal{N}$};
\node[below] at (0.3,-3.7) {$S^1\times \mathcal {M}^{d-1}$};
 \node[below] at (6.8,-3.7) {$\mathcal {M}^{d-1}$} ;
\draw[thick,dashed] (0,-0.5) to (0,-1.18);
\node[circle,draw=red,fill=red,scale=0.3] at (0,-0.5) {}; 
  \begin{scope}[rotate=-90,scale=0.5,xshift = 3 cm,yshift=-0.4cm]
    \shade[top color=red!40, bottom color=red!10]  
    (-0.5,0) to[out=140,in=-130] (0.5,2)
   to[out=60,in=200]  (2,2) 
   to[out=-10,in=130] (4,2)
   to[out=-50,in=30] (4,-0.2)
   to[out=195, in=-16] (-0.5,0);
    \draw[thick, smooth cycle,tension=.7] plot coordinates{(-0.5,0) (0.5,2) (2,2) (4,2) (4,-0.2)};
    \coordinate (A) at (1,1);
    \draw[thick, fill=white] (A) arc(140:40:1) (A) arc(-140:-20:1) (A) arc(-140:-160:1) (A) arc(140:40:1);
      \draw[ thick, white] (1.07,1) --(2.47,1);
      
    \end{scope}
   \begin{scope}[xshift = 1.25 in]
    \node[] at (0,-2) {$\rightarrow$}; 
      \end{scope}
      
        \begin{scope}[xshift = 2.5 in]
    
  \begin{scope}[rotate=-90,scale=0.5,xshift = 3 cm,yshift=-0.4cm]
    \shade[top color=red!40, bottom color=red!10]  
    (-0.5,0) to[out=140,in=-130] (0.5,2)
   to[out=60,in=200]  (2,2) 
   to[out=-10,in=130] (4,2)
   to[out=-50,in=30] (4,-0.2)
   to[out=195, in=-16] (-0.5,0);
    \draw[thick, smooth cycle,tension=.7] plot coordinates{(-0.5,0) (0.5,2) (2,2) (4,2) (4,-0.2)};
    \coordinate (A) at (1,1);
    \draw[thick, fill=white] (A) arc(140:40:1) (A) arc(-140:-20:1) (A) arc(-140:-160:1) (A) arc(140:40:1);
      \draw[ thick, white] (1.07,1) --(2.47,1);
   
    \end{scope}
    \end{scope}

  \end{tikzpicture}
\caption{Non-invertible twisted compactification: insert a non-invertible symmetry defect $\mathcal N$ at a point on $S^1$ extending along $\mathcal M^{d-1}$ and then compactify on $S^1$.}
\label{fig:nontwistcompact}
\end{center}
\end{figure}

To make the topological interface $\mathcal N$ a topological defect, we need to choose a self-dual point on the conformal manifold $\mathsf S\cdot x=x$ and let the topological manipulation cancel the change of global structures $\sigma\mathsf S\cdot\rho=\rho$. In this case $\mathcal N$ becomes a symmetry defect and generically it's a \emph{non-invertible} self-duality defect.

The study of symmetry constraints on infrared behaviors of QFTs has a long history. In this paper, instead of constraining RG flows we use non-invertible symmetry to generate entirely \emph{new} RG flows. This is done by \emph{non-invertible twisted compactification} \cite{Kaidi:2022uux}.

Let's take $\mathcal T_\rho(x)$ on $S^1\times \mathcal M^{d-1}$. If we turn on RG flow and study the far-infrared phase, we're looking at the theory at a very large scale. If we further impose the condition that the radius of $S^1$ is small compared to $\mathcal M^{d-1}$ then the IR theory will look effectively $d-1$ dimensional and we get an $d-1$ dimensional theory living on $\mathcal M^{d-1}$. This precedure of getting a lower dimensional theory from a higher dimensional theory is called \emph{compactification}. Instead of compactifying directly we can add a twist and perform twisted compactification. Recall that the ordinary twisted compactification by a symmetry element $g\in G$ is to impose $g$-twisted boundary condition on $S^1$:
\be
\mathcal O(\theta+2\pi)=g\cdot\mathcal O(\theta)
\ee
where $\theta$ parametrizes $S^1$ and $\mathcal O$ is an arbitrary operator. In modern language, this amounts to inserting the codimension-1 operator $U_g(\mathcal M^{d-1})$ at $x\in S^1$ extending along remaining directions. Now the definition of non-invertible twisted compactification comes naturally: we insert a non-invertible symmetry defect $\mathcal N$ at a point on $S^1$ extending along remaining directions and then compactify the theory, as depicted in Fig. \ref{fig:nontwistcompact}. Non-invertible twisted compactification  has been studied only for $4d$ $\mathcal N=4$ SYM in \cite{Kaidi:2022uux} (and it was studied earlier in \cite{Ganor:2008hd,Ganor:2010md,Ganor:2012mu} known as S-duality twisted compactification).

A richer set of theories, which is also constrained enough to be well studied, are class $\mathcal S$ theories. Originally introduced in \cite{Gaiotto:2009we}, class $\mathcal S$ theory is a class of 4d $\mathcal N=2$ SCFTs obtained by compactifying the 6d $\mathcal N=(2,0)$ SCFT on a punctured Riemann surface $\Sigma_{g,n}$. The theory has a duality group as the mapping class group of underlying surface $\mathrm{MCG}(\Sigma_{g,n})$ and it can be used to produce non-invertible symmetries \cite{Bashmakov:2022jtl,Bashmakov:2022uek,Antinucci:2022cdi}. Thus in principle we can indeed study compactification of class $\mathcal S$ theories on $S^1$ with a non-invertible twist.

For circle compactification of class $\mathcal S$ theory, it's known that the low energy effective 3d theory is $\mathcal N=4$ sigma model whose target space is Hitchin moduli space \cite{Gaiotto:2009hg}. The main result of this paper is that non-invertible twisted compactification of class $\mathcal S$ theory yields 3d $\mathcal N=4$ sigma model whose target space is a $(B,B,B)$ brane of Hitchin moduli space. The $(B,B,B)$ brane is the fixed point set of a finite subgroup of mapping class group on Hitchin moduli space. We will give a description of the $(B,B,B)$ branes as an affine variety. While $(B,B,B)$ branes of Hitchin moduli space arising from finite groups acting on Riemann surface have been discovered in the literature \cite{Heller:2016drq}, such a description has never appeared in mathematical literature.  We will calculate concrete examples of these $(B,B,B)$ branes for class $\mathcal S$ theory of type $A_1$ and genus 2. Our calculation shows that the $(B,B,B)$ branes have both very rich geometric structures and very rich algebraic structures.  

This paper is organized as follows. In Section \ref{section2} we illustrate the construction of non-invertible symmetries of class $\mathcal S$ theories. In Section \ref{section3.1} we review compactification of class $\mathcal S$ theories on $S^1$ with no twists. In Section \ref{section3.2} we study non-invertible twisted compactification and show that the resulting 3d theory is $\mathcal N=4$ sigma model whose target space is a $(B,B,B)$ brane of Hitchin moduli space. In Section \ref{section4} we describe these $(B,B,B)$ branes as affine varieties. In Section \ref{section4.1} we introduce a loop coordinate system for Hitchin moduli space to describe it as an affine variety. In Section \ref{section4.2} we calculate the $(B,B,B)$ branes for type $A_1$, genus 2 class $\mathcal S$ theory.
\section{Non-invertible symmetries of class $\mathcal S$ theories}\label{section2}
Non-invertible symmetries of class $\mathcal S$ theories has been studied in the literature \cite{Bashmakov:2022jtl,Bashmakov:2022uek,Antinucci:2022cdi}. In this section we summarize known results and add more details.

Quantum field theories with identical local dynamics could admit different global structures. Consider class $\mathcal S$ theories of type $A_{N-1}$, obtained by compactifying the 6d $(2,0)$ theory of type $A_{N-1}$ on a Riemann surface without punctures, which we denote by $\mathcal T[A_{N-1},\Sigma_{g,0}]$. As pointed out in \cite{Tachikawa:2013hya,Gukov:2020btk}, $\mathcal T[A_{N-1},\Sigma_{g,0}]$ has a discrete choice of global structures originating from 6d.

From 7d perspective, the 6d $(2,0)$ theory is a relative QFT, which means it lives on the  boundary of a 7d TQFT $\mathcal M_6=\partial\mathcal M_7$. As a consequence it dosen't have a partition function but only have partition vectors valued in the Hilbert space of the 7d TQFT. To specify a partition vector we must study boundary conditions of the 7d TQFT. 

The 7d TQFT has discrete 3-fluxes valued in $Z(A_{N-1})=\mathbb Z_N$, where $Z(A_{N-1})$ is the center of the simply-connected Lie group of type $A_{N-1}$. Denote a flux on the boundary by $a\in H^3(\mathcal M_6,\mathbb Z_N)$ and denote by $\Phi(a)$ the topological operator of $a$. We should think of the 6d $(2,0)$ theory as living on a spatial slice $\mathcal M_6$ of the 7d TQFT and the topological operators $\Phi(a)$ acting on the Hilbert space of the spatial slice. $\Phi(a)$ satisfies the following commutation relation:
\be\label{commurelation}
\Phi(a)\Phi(b)=\mathrm e^{i\langle a,b\rangle} \Phi(b)\Phi(a)\quad a,b\in H^3(\mathcal M_6,\mathbb Z_N)
\ee
where $\langle -,-\rangle$ is a natural pairing $H^3(\mathcal M_6,\mathbb Z_N)\times H^3(\mathcal M_6,\mathbb Z_N)\to \mathbb R/2\pi\mathbb Z$:
\be\label{pairAN}
(a,b)\mapsto {\frac{2\pi }{N}\int a\cup b}\quad a,b\in H^3(\mathcal M_6,\mathbb Z_N)
\ee 

We cannot simultaneously fix the values $a,b$ if they have a non-trivial pairing. Instead we can do as follows. Since the pairing \eqref{pairAN} is antisymmetric and non-degenerate, it is actually symplectic and we can specify two Lagrangian (maximally isotropic) lattices $\mathcal L,\mathcal L^\bot\subset H^3(\mathcal M_6,\mathbb Z_N)$ such that
\be
\mathcal L\oplus\mathcal L^\bot=H^3(\mathcal M^6,\mathbb Z_N)
\ee
$\mathcal L$ is a maximal lattice in which elements have trivial pairings and it specifies the fluxes for which we impose Dirichlet boundary conditions (i.e. setting to zero). By the canonical commutation relation \eqref{commurelation} we must impose Neuman boundary condition on the complement $\mathcal L^\bot$. We say such a split $(\mathcal L,\mathcal L^\bot)$ defines a \emph{global structure} for the 6d $(2,0)$ theory. The pair $(\mathcal L,\mathcal L^\bot)$ is also called a \emph{polarization pair} in \cite{Lawrie:2023tdz}. As we impose Dirichlet boundary condition for $\mathcal L$, the topological operator $\Phi(a)\quad a\in\mathcal L$ has trivial action. Instead $\mathcal L^\bot$ is imposed Neuman boundary condition and topological operators of $\mathcal L^\bot$ act non-trivially on the partition vector. Therefore we see that $\mathcal L^\bot$ determines the \emph{generalized global symmetries} of the theory. Denote a partition vector by $Z_{(\mathcal  L,\mathcal L^\bot)}[b]$, where $b\in\mathcal L^\bot$, then:
\be
\Phi(b')Z_{(\mathcal L,\mathcal L^\bot)}[b]=Z_{(\mathcal L,\mathcal L^\bot)}[b+b']\quad \Phi(a)Z_{(\mathcal L,\mathcal L^\bot)}[b]=\mathrm e^{i\langle a,b\rangle}Z_{(\mathcal L,\mathcal L^\bot)}[b]
\ee
as expected. Thus the partition vectors $Z_{(\mathcal  L,\mathcal L^\bot)}[b]\quad b\in\mathcal L^\bot$ span a complete basis in the Hilbert space of the 7d TQFT.

A \emph{topological manipulation} is defined to be a map from a possible global structure $(\mathcal L,\mathcal L^\bot)$ to another. From the 7d perspective it's an isomorphism of topological boundary conditions on $\mathcal M_6=\partial \mathcal M_7$. Isomorphisms of topological boundary conditions form a groupoid, named \emph{orbifold groupoid} in \cite{Gaiotto:2020iye}. In 6d language, it can be rephrased as consisting of gauging a $q$-form global symmetry and stacking an SPT phase. For example the map $(\mathcal L,\mathcal L^\bot)\mapsto (\mathcal L^\bot,\mathcal L)$ is the familiar “gauging a $q$-form global symmetry" \cite{Tachikawa:2017gyf,Bhardwaj:2017xup}:
\be
Z_{(\mathcal L^\bot,\mathcal L)}[a]=\sum_{b\in\mathcal L^\bot}{\mathrm{e}^{i\langle a,b\rangle}Z_{(\mathcal L,\mathcal L^\bot)}[b]}\quad a\in\mathcal L
\ee

How can we represent a topological manipulation? We should first represent the global structure in a convenient way. Let $n$ be such that $H^3(\mathcal M^6,\mathbb Z_N)=(\mathbb Z_N)^{2n}$. Suitably choosing a basis $\{v_i\}$ for $\mathcal L$ and $\{v_{n+j}\}$ for $\mathcal L^\bot$, we can bring the symplectic pairing \eqref{pairAN} into the canonical form:
\be\label{canopairing}
\langle v_i,v_j\rangle=\left(\begin{matrix}0&-\mathds1_{n\times n}\\\mathds 1_{n\times n}&0\\\end{matrix}\right)
\ee

We may say a basis $\{v_1,\dots,v_n,v_{n+1},\dots,v_{2n}\}$ satisfying \eqref{canopairing} defines a global structure. Then another basis $\{v'_i\}$ can be represented by a matrix $\mathscr M$ by expanding the vectors $v'_i$ under the basis $\{v_i\}$:
\be
(v'_1\cdots v'_{2g})=(v_1 \cdots v_{2g})\mathscr M
\ee
 The condition that $\{v'_i\}$ also satisfies \eqref{canopairing} translates into the requirement that $\mathscr M$ is a symplectic matrix:
\be
\mathscr M^T\mathfrak J\mathscr M=\mathfrak J
\ee
where $\mathfrak J=\left(\begin{matrix}0&-\mathds1_{n\times n}\\\mathds 1_{n\times n}&0\\\end{matrix}\right)$ is the canonical symplectic matrix. Thus $\mathscr M\in Sp(2n,\mathbb Z_N)$. However note that the correspondence between $\mathscr M$ and the global structure $(\mathcal L,\mathcal L^\bot)$ is not one-to-one: we may redefine the basis of $\mathcal L$ so that it also satisfies \eqref{canopairing}. Denote by $Q\in GL(n,\mathbb Z_N)$ the matrix that changes the basis $\{v'_1,\dots,v'_n\}$, then we must change the basis $\{v'_{n+1},\dots,v'_{2n}\}$ of $\mathcal L^\bot$ accordingly by $Q'$ to maintain \eqref{canopairing}. A straight computation gives:
\be
\sum_i Q_{im}Q'_{in}=\delta_{mn}
\ee   
Thus we have $Q'=(Q^T)^{-1}$. Acting on $\mathscr M$ by $\mathrm{diag}(Q,Q')$ doesn't correspond to a change of global structure. Finally we see a global structure is represented by an element in
\be
\frac{Sp(2n,\mathbb Z_N)}{GL(n,\mathbb Z_N)}
\ee
As a consequence the number of global structures is:
\be
d=\frac{|Sp(2n,\mathbb Z_N)|}{|GL(n,\mathbb Z_N)|}
\ee 

A topological manipulation is a linear transformation of $\{v'_i\}$ preserving \eqref{canopairing}. Thus we see clearly that a topological manipulation is an element of the symplectic group $Sp(2n,\mathbb Z_N)$, acting on $\mathscr M$ on the right. Note also a symplectic transformation by $\mathrm{diag}(Q,Q')$ doesn't correspond to a change of global structure. We draw the conclusion that topological manipulations of 6d $(2,0)$ theory of type $A_{N-1}$ form the group $Sp(2n,\mathbb Z_N)/GL(n,\mathbb Z_N)$.

Upon compactification to 4d we have $\mathcal M_6=\mathcal M_4\times\Sigma_{g,0}$ and the cohomology group splits by K\"{u}nneth formula:
\be
\begin{gathered}
H^3(\mathcal M_6,\mathbb Z_N)\simeq ( H^2(\mathcal M_4,\mathbb Z_N)\otimes H^1(\Sigma_{g,0},\mathbb Z_N))\\ 
\oplus(H^1(\mathcal M_4,\mathbb Z_N)\otimes H^2(\Sigma_{g,0},\mathbb Z_N))\oplus (H^3(\mathcal M_4,\mathbb Z_N)\otimes H^0(\Sigma_{g,0},\mathbb Z_N))
\end{gathered}
\ee

In $H^1(\mathcal M_4,\mathbb Z_N)\otimes H^2(\Sigma_{g,0},\mathbb Z_N)$ and $H^3(\mathcal M_4,\mathbb Z_N)\otimes H^0(\Sigma_{g,0},\mathbb Z_N)$ the pairing \eqref{pairAN} is trivial and thus we need only to consider the first summand $H^2(\mathcal M_4,\mathbb Z_N)\otimes H^1(\Sigma_{g,0},\mathbb Z_N)$. Split first $H^1(\Sigma_{g,0},\mathbb Z_N)$ into two Lagrangian sublattices $L,L^\bot\subset H^1(\Sigma_{g,0},\mathbb Z_N)$, then 
\be
H^2(\mathcal M_4,\mathbb Z_N)\otimes H^1(\Sigma_{g,0},\mathbb Z_N)=(H^2(\mathcal M_4,\mathbb Z_N)\otimes L)\oplus(H^2(\mathcal M_4,\mathbb Z_N)\otimes L^\bot)
\ee
The first and second term on the RHS correspond to “electric" and “magnetic" fluxes through 2-cycles of $\mathcal M^4$ respectively. As a consequence $L$ and $L^\bot$ are charges of “Wilson" and “'t-Hooft" lines respectively. We refer to $H^1(\Sigma_{g,0},\mathbb Z_N)$ as the \emph{line lattice} of class $\mathcal S$ theories $\mathcal T[A_{N-1},\Sigma_{g,0}]$, as it specifies charges of line operators of the theory. The symplectic pairing in $H^1(\Sigma_{g,0},\mathbb Z_N)$ is the \emph{Dirac pairing} for line operators. Only line operators with trivial Dirac pairing could exist simultaneously in the theory. Thus $L$ is the \emph{spectrum of allowed line operators} and $L^\bot$ defines \emph{1-form symmetries} of the theory \cite{Aharony:2013hda}. The pair $(L,L^\bot)$ is called a \emph{global structure} of class $\mathcal S$ theories.

We can represent a global structure $(L,L^\bot)$ similarly as we did in 6d. Note that $H^1(\Sigma_{g,0},\mathbb Z_N)\simeq (\mathbb Z_N)^{2g}$. We choose a basis $\{u_1,\dots,u_{2g}\}$ of $H^1(\Sigma_{g,0},\mathbb Z_N)$ such that 
\be\label{canopairing4d}
\langle u_i,u_j\rangle=\left(\begin{matrix}0&-\mathds1_{g\times g}\\\mathds 1_{g\times g}&0\\\end{matrix}\right)
\ee 
The first $g$ and the last $g$ vectors span a basis for $L$ and $L^\bot$, respectively. Such a basis $\{u_i\}$ defines a global structure and another basis $\{u'_i\}$ can be represented by a matrix $\mathscr M$ by expanding it under $\{u_i\}$:
\be
(u'_1\cdots u'_{2g})=(u_1\cdots u_{2g})\mathscr M
\ee
The requirement that $\{u'_i\}$ also satisfies \eqref{canopairing4d} is equivalent to requiring $\mathscr M$ to be a symplectic matrix $\mathscr M\in Sp(2g,\mathbb Z_N)$:
\be
\mathscr M^T\mathfrak J\mathscr M=\mathfrak J
\ee
where $\mathfrak J=\left(\begin{matrix}0&-\mathds1_{g\times g}\\\mathds 1_{g\times g}&0\\\end{matrix}\right)$ is the canonical symplectic matrix.\footnote{By an abuse of language we use $\mathscr M$ and $\mathfrak J$ to denote corresponding quantities both in 4d and 6d case.} A redefinition of the basis for $L$ doesn't correspond to a change of global structure. These transformations are implemented by $\mathrm{diag}(Q,(Q^T)^{-1})$ acting on $\mathscr M$ on the right where $Q\in GL(g,\mathbb Z_N)$. Thus the number of global structures of class $\mathcal S$ theories is:
\be
d=\frac{|Sp(2g,\mathbb Z_N)|}{|GL(g,\mathbb Z_N)|}
\ee
For simplicity we will denote a global structure by $\mathscr M\in Sp(2g,\mathbb Z_N)$, ignoring the redundency.

A topological manipulation is a linear transformation of $\{u'_i\}$ preserving \eqref{canopairing4d}. Thus a topological manipulation is also represented by a symplectic matrix $G\in Sp(2g,\mathbb Z_N)$, acting on $\mathscr M$ on the right:
\be\label{actiontopomani}
G:\mathscr M\mapsto \mathscr M G \quad G\in Sp(2g,\mathbb Z_N)
\ee
Note that $G$ is also defined up to $\mathrm{diag}(Q,(Q^T)^{-1})$ but we will ignore the redundency and denote a topological manipulation simply by $G\in Sp(2g,\mathbb Z_N)$.

Having worked out the action of topological manipulations on $\mathscr M$, how does the duality group act on the global structures? We shall first review the conformal manifolds and duality groups of class $\mathcal S$ theories.

The conformal manifold of a class $\mathcal S$ theory $\mathcal T[A_{N-1},\Sigma_{g,0}]$ is the moduli space of complex structures on $\Sigma_{g,0}$, the Teichmuller space $\mathrm{Teich}(\Sigma_{g,0})$. For convenience we will express the complex strcture as the \emph{period matrix} $\Omega$ of $\Sigma_{g,0}$. Indeed by Torelli theorem the complex structure is completely determined by $\Omega$.

$\Omega$ is constructed as follows. Let $H_1(\Sigma_{g,0},\mathbb Z)$ be the first homology group of $\Sigma_{g,0}$. There is a symplectic intersection pairing in $H_1(\Sigma_{g,0},\mathbb Z)\simeq \mathbb Z^{2g}$ and we can decompose $H_1(\Sigma_{g,0},\mathbb Z)$ into $\mathsf A$-cycles and $\mathsf B$-cycles according to the symplectic pairing in the standard way. That is, choose a basis $\{\mathsf A_I,\mathsf B_J\}\quad I,J=1,\dots,g$ of $H_1(\Sigma_{g,0},\mathbb Z)$ such that:
\be
\langle \mathsf A_I,\mathsf A_J\rangle=0\quad\langle \mathsf B_I,\mathsf B_J\rangle=0\quad \langle \mathsf A_I,\mathsf B_J\rangle=\delta_{IJ}
\ee 
Denote the basis by $\{v_i\}=\{\mathsf B_1,\dots,\mathsf B_g,\mathsf A_1,\dots,\mathsf A_g\}$ then $\{v_i\}$ brings the intersection pairing to the canonical form:
\be\label{canopairingAB}
\langle v_i,v_j\rangle=\left(\begin{matrix}0&-\mathds1_{g\times g}\\\mathds 1_{g\times g}&0\\\end{matrix}\right)
\ee
Let $H^{(1,0)}(\Sigma_{g,0})$ be the space of holomorphic one-forms on the Riemann surface $\Sigma_{g,0}$. It has complex dimension $g$: $H^{(1,0)}(\Sigma_{g,0})\simeq \mathbb C^g$. We can choose a basis $\{\omega_I\}$ of $H^{(1,0)}(\Sigma_{g,0})$ which is dual to the $\mathsf A$-cycles:
\be
\oint_{\mathsf A_I}\omega_J=\delta_{IJ}
\ee
Then the integration of $\{\omega_J\}$ over $\mathsf B$-cycles is called the \emph{period matrix} of the Riemann surface $\Sigma_{g,0}$:
\be
\Omega_{IJ}:=\oint_{\mathsf B_I}\omega_J
\ee

The period matrix $\Omega=\Omega_{IJ}$ is a $g\times g$ complex matrix and it has the property that it is symmetric and its imaginary part has positive determinant:
\be
\Omega_{IJ}=\Omega_{JI} \quad\mathrm{det}(\mathrm{Im}(\Omega_{IJ}))>0
\ee
We can view $\Omega_{IJ}$ as encoding the coupling constants of class $\mathcal S$ theories and it parametrizes the conformal manifold.

The duality group of class $\mathcal S$ theories is the mapping class group $\mathrm{MCG}(\Sigma_{g,0})$ of the underlying Riemann surface $\Sigma_{g,0}$. There is a short exact sequence for $\mathrm{MCG}(\Sigma_{g,0})$:
\be
\mathrm{Tor}\to\mathrm{MCG}(\Sigma_{g,0})\to Sp(2g,\mathbb Z)
\ee
where $\mathrm{Tor}$ is the Torelli group and it acts trivially on $H_1(\Sigma_{g,0},\mathbb Z)$. Thus the action of mapping class group is equivalent to the action of $Sp(2g,\mathbb Z)$ and we will simply refer to the duality group as $Sp(2g,\mathbb Z)$. Note that more precisely the duality group is mapping class group $\mathrm{MCG}(\Sigma_g)$ but not $Sp(2g,\mathbb Z)$. But in this section we are dealing with the action of duality group on global structures, for this purpose it's enough to regard the duality group as $Sp(2g,\mathbb Z)$. We will denote an element of duality group by
\be
F=\left(\begin{matrix}A&B\\C&D\\\end{matrix}\right)\in Sp(2g,\mathbb Z)
\ee
The block matrices $A,B,C,D$ satisfy:
\be
AB^T=BA^T\quad CD^T=DC^T\quad AD^T-BC^T=\mathds 1_{g\times g}
\ee 

The action of duality group $Sp(2g,\mathbb Z)$ on $H_1(\Sigma_{g,0},\mathbb Z)$ is standard:
\be
F:\left(\begin{matrix}\mathsf B\\\mathsf A\\\end{matrix}\right)\to \left(\begin{matrix}A&B\\C&D\\\end{matrix}\right)\left(\begin{matrix}\mathsf B\\\mathsf A\\\end{matrix}\right)
\ee
As a consequence the action of $F$ on $\Omega$ is the linear fractional transformaation:
\be\label{dualityaction}
F:\Omega\to (A\Omega+B)(C\Omega+D)^{-1}
\ee

For example, when $g=1$ our theory is 4d $\mathcal N=4$ SYM and the period matrix $\Omega$ reduces to a complex number $\tau$, which is identified with the complex coupling constant of 4d $\mathcal N=4$ SYM. The duality group reduces to $SL(2,\mathbb Z)$ and \eqref{dualityaction} becomes the familiar $SL(2,\mathbb Z)$ duality of 4d $\mathcal N=4$ SYM:
\be
\tau\to\frac{a\tau+b}{c\tau+d} \quad\left(\begin{matrix}a&&b\\c&&d\\\end{matrix}\right)\in SL(2,\mathbb Z)
\ee  
It's generated by $T$-transformation and $S$-transformation. When $g=2$ the period matrix takes the form:
\be
\Omega=\left(\begin{matrix}\tau_1&\tau_2\\\tau_2&\tau_3\\\end{matrix}\right)
\ee
where $\tau_2\to 0$ limit corresponds to the degeneration limit of the genus-$2$ Riemann surface: it splits into two tori with complex moduli $\tau_1$ and $\tau_3$. The duality group becomes $Sp(4,\mathbb Z)$. It has 4 generators and we may represent them as follows:
\be
Sp(4,\mathbb Z)=\langle T_1,T_2,S_{12},W\rangle
\ee
Explicitly the four generators are:
\be
T_1=\left(\begin{matrix}1&0&1&0\\0&1&0&0\\0&0&1&0\\0&0&0&1\\\end{matrix}\right)
\quad T_2=\left(\begin{matrix}1&0&0&0\\0&1&0&1\\0&0&1&0\\0&0&0&1\\\end{matrix}\right)
\quad S_{12}=\left(\begin{matrix}0&0&\,1&0\\0&0&0&\,1\\-1&0&0&0\\0&-1&0&0\\\end{matrix}\right)
\quad W=\left(\begin{matrix}1&0&0&1\\0&1&1&0\\0&0&1&0\\0&0&0&1\\\end{matrix}\right)
\ee
They act on $\mathsf A$-cycles and $\mathsf B$-cycles as:
\be
\begin{aligned}
T_1\left(\begin{matrix}\mathsf B_1\\\mathsf B_2\\\mathsf A_1\\\mathsf A_2\\\end{matrix}\right)&=\left(\begin{matrix}\mathsf B_1+\mathsf A_1\\\mathsf B_2\\\mathsf A_1\\\mathsf A_2\\\end{matrix}\right)
\quad T_2\left(\begin{matrix}\mathsf B_1\\\mathsf B_2\\\mathsf A_1\\\mathsf A_2\\\end{matrix}\right)&=\left(\begin{matrix}\mathsf B_1\\\mathsf B_2+\mathsf A_2\\\mathsf A_1\\\mathsf A_2\\\end{matrix}\right)\\
S_{12}\left(\begin{matrix}\mathsf B_1\\\mathsf B_2\\\mathsf A_1\\\mathsf A_2\\\end{matrix}\right)&=\left(\begin{matrix}\mathsf A_1\\\mathsf A_2\\-\mathsf B_1\\-\mathsf B_2\\\end{matrix}\right)
\quad\quad\,\,\, W\left(\begin{matrix}\mathsf B_1\\\mathsf B_2\\\mathsf A_1\\\mathsf A_2\\\end{matrix}\right)&=\left(\begin{matrix}\mathsf B_1+\mathsf A_2\\\mathsf B_2+\mathsf A_1\\\mathsf A_1\\\mathsf A_2\\\end{matrix}\right)
\end{aligned}
\ee
We recognize $T_1$ and $T_2$ as the separate $T$-transformations on the two sub-tori. $S_{12}$ is a simultaneous $S$-transformation on both tori. $W$ has no analog to $T$- or $S$-transformations and it's an intrinsic genus-2 transformation.

Now we are ready to determine the action of duality group on global structures. Having chosen a standard basis $\{u_i\}$ of $H^1(\Sigma_{g,0},\mathbb Z_N)$ satisfying \eqref{canopairing4d}, it also determines a dual basis $\{v_i\}$ of $H_1(\Sigma_{g,0},\mathbb Z_N)$ satisfying \eqref{canopairingAB}. We may say the basis $\{v_i\}$ determines $\mathsf B$-cycles and $\mathsf A$-cycles of the Riemann surface. The condition that $\{u_i\}$ and $\{v_i\}$ are dual to each other can be rephrased in matrix language:
\be\label{dualpairing}
\left(\begin{matrix}v_1\\\vdots\\v_{2g}\\\end{matrix}\right)\left(\begin{matrix}u_1&\cdots&u_{2g}\end{matrix}\right)=\mathds 1_{2g\times 2g}
\ee
where the multiplication $(v_i\cdot u_j)$ is defined to be the pairing $u_j(v_i)$. Identifying $\{v_i\}$ as $\mathsf B$-cycles and $\mathsf A$-cycles, we see the action of $F$ on $\{v_i\}$ is:
\be
\left(\begin{matrix}v_1\\\vdots\\v_{2g}\\\end{matrix}\right)\to F \left(\begin{matrix}v_1\\\vdots\\v_{2g}\\\end{matrix}\right)
\ee
We must act on $\{u_i\}$ accordingly by $F'$ to maintain \eqref{dualpairing}:
\be
F\left(\begin{matrix}v_1\\\vdots\\v_{2g}\\\end{matrix}\right)\left(\begin{matrix}u_1&\cdots&u_{2g}\end{matrix}\right)F'=\mathds 1_{2g\times 2g}
\ee
Thus we see clearly that $F'=F^{-1}$. The duality action amounts to a change of the standard basis. The global structure is parametrized by the expansion $\mathscr M$ under the standard basis. So the duality changes $\mathscr M$ by $F''$ which satisfies:
\be
(u_1\cdots u_{2g})\mathscr M=(u_1\cdots u_{2g})F'F''\mathscr M
\ee
Finally we have $F''=(F')^{-1}=F$ and the duality group acts on $\mathscr M$ by
\be
\mathscr M\to F\mathscr M\quad F\in Sp(2g,\mathbb Z)\footnote{Note that while $\mathscr M$ is valued in $\mathbb Z_N$, $F$ is valued in $\mathbb Z$. So actually we should first mod $F$ by $N$ to map it to $Sp(2g,\mathbb Z_N)$ and then multiply it on $\mathscr M$.}
\ee

Combining it with \eqref{actiontopomani}, we finally work out the full transformation of global structures:
\be\label{transglobalstructure}
\mathscr M\to F\mathscr M G
\ee
We shall give some remarks on \eqref{transglobalstructure}:
\begin{enumerate}
\item The duality $F$ acts on $\mathscr M$ on the \emph{left} while the topological manipulation $G$ acts on the \emph{right}. This assures, \emph{a priori}, that the actions of duality and topological manipulation commute with each other.
\item Given an arbitrary duality $F$, we are able to find an appropriate topological manipulation $G$ to undo the change of the global structure.
\end{enumerate}

\begin{figure}[tbp]
\begin{center}
\[\hspace*{-0.8 cm}\begin{tikzpicture}[baseline=19,scale=0.8]
\draw[  thick] (0,-0.2)--(0,3.2);
\draw[  thick] (3,-0.2)--(3,3.2);
  
          \shade[line width=2pt, top color=red,opacity=0.4] 
    (0,0) to [out=90, in=-90]  (0,3)
    to [out=0,in=180] (3,3)
    to [out = -90, in =90] (3,0)
    to [out=190, in =0]  (0,0);
    
\node[left] at (-0.7,1.5) {$\mathcal T_{\mathscr M}(\Omega)$};
    \node at (1.5,1.5) {$\mathcal T_{F\mathscr M}(F(\Omega))$};
    \node[right] at (3.7,1.5) {$\mathcal T_{F\mathscr M G}(F(\Omega))$};
      \node[left] at (0,-0) {$F$};
          \node[right] at (3,-0) {$G$};  
\end{tikzpicture}
\hspace{0.35 in}\Rightarrow\hspace{0.35 in}
\begin{tikzpicture}[baseline=19,scale=0.8]
\draw[red, thick] (0,-0.2)--(0,3.2);
  \node[left] at (-0.7,1.5) {$\mathcal T_{\mathscr M}(\Omega)$};
    \node[right] at (+0.7,1.5) {$\mathcal T_{F\mathscr MG}(F(\Omega))$};
      \node[left] at (0,0) {$\mathcal N$};
\end{tikzpicture}
\]
\caption{Construction of non-invertible symmetries of class $\mathcal S$ theories.}
\label{fig:NclassS}
\end{center}
\end{figure}
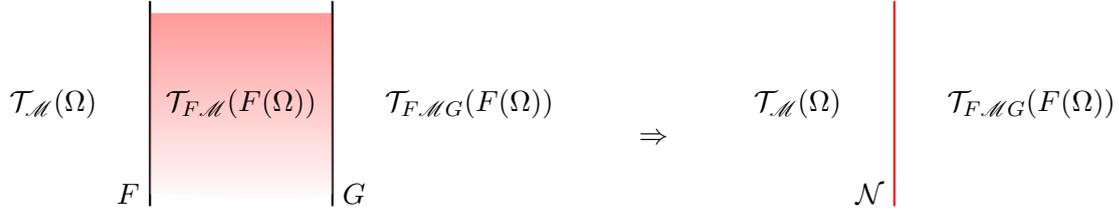

Now let's recall the construction of \emph{non-invertible self-duality defect}. Denote class $\mathcal S$ theories of genus $g$ and type $A_{N-1}$ by $\mathcal T_{\mathscr M}(\Omega)$, where $\Omega$ is the period matrix and $\mathscr M$ labels a global strcture. The duality $F$ acts on $\mathcal T_{\mathscr M}(\Omega)$ by $\mathcal T_{\mathscr M}(\Omega)\mapsto\mathcal T_{F\mathscr M}(F(\Omega))$. To map the system back to itself we should dress the duality by a topological manipulation $G$ and the theory becomes $\mathcal T_{F\mathscr MG}(F(\Omega))$, as depicted in Fig. \ref{fig:NclassS}. To make $\mathcal N=GF$ a topological defect but not a topological interface we should have:
\be\label{conditionforN}
F\mathscr M G=\mathscr M\quad F(\Omega)=\Omega
\ee
As remarked above, we are always able to find $G$ satisfying the first equation. So we only need to focus on solutions of the second equation. For $g=1$ our theory is 4d $\mathcal N=4$ SYM and the solutions are $\tau=i$, $e^{\frac{2\pi i}{3}}$, as studied extensively in the literature. For higher genus, to the author's knowledge, it's been solved only for $g\leq 5$ \cite{Gottschling1961,Gottschling1961b,Gottschling1967,KURIBAYASHI1990277,Kuribayashi1986OnAG}. When $g=2$ the duality group is $Sp(4,\mathbb Z)$ and the solutions are listed in Table \ref{tab:fixedpoints}.\footnote{The same result can also be found in \cite{Bashmakov:2022uek,Nilles:2021glx}. Our notations coincide with \cite{Bashmakov:2022uek}.} The generators in Table \ref{tab:fixedpoints} are explicitly collected in Appendix \ref{appendixA}.

\begin{table}[tbp]
\centering

\begin{tabular}{|c|cccc|}
\hline Complex dimension&Order&Subgroup&Generators&$\Omega$\\
\hline 0&10&$\mathbb Z_{10}$&$\phi$&$\left(\begin{matrix}\varepsilon&\varepsilon+\varepsilon^{-2}\\\varepsilon+\varepsilon^{-2}&-\varepsilon^{-1}\\\end{matrix}\right)$\\
&24&$(\mathbb Z_2\times\mathbb Z_6)\rtimes\mathbb Z_2$&$M_1,M_2,M_3$&$\frac{i}{\sqrt{3}}\left(\begin{matrix}2&1\\1&2\end{matrix}\right)$\\
&&$\mathbb Z_{12}\times \mathbb Z_2$&$C,M_4$&$\left(\begin{matrix}\rho&0\\0&i\\\end{matrix}\right)$\\
&32&$(\mathbb Z_4\times\mathbb Z_4)\rtimes\mathbb Z_2$&$M_5,M_6,M_7$&$\left(\begin{matrix}i&0\\0&i\\\end{matrix}\right)$\\
&48&$GL(2,3)$&$M_7,M_8$&$\frac{1}{3}\left(\begin{matrix}1+2\sqrt 2i&-1+\sqrt 2i\\-1+\sqrt 2i&1+2\sqrt 2i\\\end{matrix}\right)$\\
&72&$\mathbb Z_3\times(\mathbb Z_6\times Z_2)\rtimes \mathbb Z_2$&$M_7,M_9,M_{10}$&$\left(\begin{matrix}\rho&0\\0&\rho\\\end{matrix}\right)$\\
\hline 1&8&$\mathbb Z_2\times\mathbb Z_4$&$C,N_1$&$\left(\begin{matrix}i&0\\0&\tau_3\\\end{matrix}\right)$\\
&&$D_8$&$M_7,N_2$&$\left(\begin{matrix}\tau_1&\frac{1}{2}\\\frac{1}{2}&\tau_1\\\end{matrix}\right)$\\
&&$D_8$&$M_7,N_3$&$\left(\begin{matrix}\tau_1&0\\0&\tau_1\\\end{matrix}\right)$\\
&12&$\mathbb Z_2\times\mathbb Z_6$&$C,N_4$&$\left(\begin{matrix}\rho&0\\0&\tau_3\\\end{matrix}\right)$\\
&&$D_{12}$&$N_5,N_6$&$\left(\begin{matrix}\tau_1&\frac{1}{2}\tau_1\\\frac{1}{2}\tau_1&\tau_1\\\end{matrix}\right)$\\
\hline 2&4&$\mathbb Z_2\times\mathbb Z_2$&$C,P$&$\left(\begin{matrix}\tau_1&0\\0&\tau_3\end{matrix}\right)$\\
&&$\mathbb Z_2\times\mathbb Z_2$&$C,M_7$&$\left(\begin{matrix}\tau_1&\tau_2\\\tau_2&\tau_1\\\end{matrix}\right)$\\
\hline
\end{tabular}
\caption{Fixed points of $Sp(4,\mathbb Z)$, where $\varepsilon:=e^{2\pi i/5}$, $\rho:=e^{2\pi i/3}$ and $\tau_1,\tau_3$ are free variables.}
\label{tab:fixedpoints}
\end{table}

Genus-$2$ case will be our main focus and we shall give a description of Table \ref{tab:fixedpoints}.
\begin{enumerate}
\item The generator $C=-\mathds 1_{2g\times 2g}$ acts trivially on every $\Omega$. Actually $C$ is the \emph{charge conjugation} operator. Excluding this trivial case, the remaining fixed loci have complex dimension 0,1,2 respectively. The complex dimension 0 loci are isolated points, while the other two types have a continuous family of points.
\item Points at which $\Omega$ is diagonal correspond to the degeneration limits of the genus-2 surface: the surface splits into two sub-tori. Denote such points by $\left(\begin{matrix}\tau_1 &0\\0&\tau_3\end{matrix}\right)$. The generators $P$ and $CP$ are separate “charge conjugations" for each sub-tori, respectively. So they must leave these points invariant. To have some enhanced symmetry beyond $C$ and $P$, we only need to choose $\tau_1$ or $\tau_3$ to be fixed by some elements in $SL(2,\mathbb Z)$. Thus we have $\tau_1,\tau_3=i,e^{2\pi i/3}$. Or we could let $\tau_1=\tau_3$ to have the enhanced permutation symmetry of the two sub-tori. Altogether we have 7 choices and these exhaust the 7 diagonal cases in Table \ref{tab:fixedpoints}.
\item The remaining fixed points are intrsically genus 2. There're altogether 6 such cases in Table \ref{tab:fixedpoints}.
\end{enumerate}

Having chosen $(F,\Omega,\mathscr M,G)$ satisfying \eqref{conditionforN}, we may ask whether the topological defect $\mathcal N$ is non-invertible and whether $\mathcal N$ is \emph{intrinsically} non-invertible. It's determined as follows: 
\begin{enumerate}
\item[$\bullet$] If $F$ doesn't change the global structure, then $\mathcal N$ is invertible.
\item[$\bullet$] If $F$ changes the global structure, then $\mathcal N$ is non-invertible.
\begin{enumerate}
\item[$\blacktriangleright$] If there exists another $\mathscr M'$ such that $F$ doesn't change the global structure, then $\mathcal N$ is non-intrinsically non-invertible.
\item[$\blacktriangleright$] If there doesn't exist $\mathscr M'$ such that $F$ doesn't change the global structure, then $\mathcal N$ is intrinsically non-invertible.  
\end{enumerate}
\end{enumerate}
Note crucially that the correspondence between $\mathscr M$ and the global structure is not one-to-one. So by “$F$ doesn't changes the global structure" we mean “$F\mathscr M$ is equal to $\mathscr M$ up to a $\mathrm{diag}(Q,(Q^T)^{-1})$".
\section{Non-invertible twisted compactification on $S^1$}\label{section3}
\subsection{Straight compactification on $S^1$ and Hitchin moduli space}\label{section3.1}
Before studying non-invertible twisted compactification of class $\mathcal S$ theories on $S^1$, we should first understand compactification of class $\mathcal S$ theories on $S^1$ with no twists. It turns out that compactification of class $\mathcal S$ theories on $S^1$ leads to 3d $\mathcal N=4$ sigma model whose target space is Hitchin moduli space \cite{Gaiotto:2009hg}.

In general, if we have a $d$-dimensional theory on $M^{d-n}\times L_n$ and compactify it on $L_n$, the moduli space of vacua of resulting $(d-n)$ dimensional theory is easy to determine: $\mathcal M_{\mathrm{vacua}}^{(d-n)}$ is the space of field configurations of the $d$ dimensional theory that satisfy equation of motion along $L_n$. The difference between $\mathcal M_{\mathrm{vacua}}^{(d-n)}$ and the moduli space of vacua of original $d$-dimensional theory is that $\mathcal M_{\mathrm{vacua}}^{(d-n)}$ doesn't require the field configuration to remain constant on $L_n$. Thus $\mathcal M_{\mathrm{vacua}}^{(d-n)}$ is a fibration over the moduli space of vacua of original $d$-dimensional theory:
\be
\mathcal M_{\mathrm{vacua}}^{(d-n)}\to\mathcal M_{\mathrm{vacua}}^{(d)}
\ee

If our theory is 4d $\mathcal N=2$ gauge theory and we compactify it on $S^1$, then the moduli space of vacua of resulting 3d theory is a fibration over 4d moduli space of vacua. The moduli space of vacua of 4d $\mathcal N=2$ theory has a Coulomb branch and a Higgs branch. At a generic point of 4d moduli space of vacua the Higgs branch is absent and there's only Coulomb branch $\mathfrak B$. Thus the moduli space of vacua of resulting 3d theory $\mathcal M$ is a fibration over 4d Coulomb branch:
\be\label{fibration3d}
\pi:\mathcal M\to\mathfrak B
\ee
The result of \cite{Gaiotto:2010okc} is that the above fibration \eqref{fibration3d} is just Seiberg-Witten fibration for 4d $\mathcal N=2$ gauge theories \cite{Seiberg:1994rs,Seiberg:1994aj,Gomez:1995rk} and the resulting 3d theory is 3d $\mathcal N=4$ sigma model with target space $\mathcal M$.

When it comes to class $\mathcal S$ theories, the above fibration is the famous Hitchin fibration and $\mathcal M$ is the Hitchin moduli space associated to the underlying Riemann surface \cite{Gaiotto:2009hg}.

Let $\Sigma_g$ be a Riemann surface of genus $g$ and we denote the canonical bundle of $\Sigma_g$ by $K$. Let $E$ be a principal $G$-bundle over $\Sigma_g$ and $A$ be a connection on $E$. The connection $A$ turns $E$ into a holomorphic $G$-bundle: indeed we can define the $\bar\partial$ operator as $\bar \partial=\mathrm d\bar z D_{\bar z}$. A Higgs field $\varphi$ is a holomorphic one-form valued in the adjoint representation: $\varphi\in\Omega^{(1,0)}(\Sigma_g,\mathrm{ad}(E)\otimes K)$. The pair $(E,\varphi)$ is called a Higgs bundle. The Hitchin moduli space $\mathcal M_{H}(G,\Sigma_g)$ consists of pairs $(A,\varphi)$ which satisfy the Hitchin equation \cite{Hitchin:1986vp}:
\begin{equation}\label{Hitchinequation}
\begin{gathered}
F_A+[\varphi,\bar\varphi]=0 \\
\bar{\partial}_A\varphi=0
\end{gathered}
\end{equation}
modulo $G$ gauge transformations. It's proved in \cite{Hitchin:1986vp} that $\mathcal M_H(G,\Sigma_g)$ is the same as the moduli space of semi-stable Higgs bundles $(E,\varphi)$ which have vanishing first Chern class. If we define a $G_{\mathbb C}$-valued connection $\mathcal A=A+i\varphi+i\bar\varphi$ then the above equation is just saying that the curvature $\mathcal F$ of the complexified connection $\mathcal A$ is zero. By Riemann-Hilbert correspondence the space of flat connections is isomorphic to the space of representations of the fundamental group. Thus we have:
\be
\mathcal M_H(G,\Sigma_g)\simeq\mathrm{Hom}(\pi_1(\Sigma_g),G_{\mathbb C})//G_{\mathbb C}
\ee
The RHS is the space of homomorphisms from $\pi_1(\Sigma_g)$ to $G_{\mathbb C}$ up to a $G_{\mathbb C}$ conjugation and it's called the $G_{\mathbb C}$-\emph{character variety}. 

The Hitchin moduli space $\mathcal M_H(G,\Sigma_g)$ has a very rich geometric structure. From the view of symplectic geometry it is an integrable system, which means there's a fibration:
\be
\pi:\mathcal M_H(G,\Sigma_g)\to \mathfrak B
\ee  
where the base $\mathfrak B$ is a complex vector space of dimension $\frac{1}{2}\mathrm{dim}\,\mathcal M_H(G,\Sigma_g)$ and the generic fiber $\pi^{-1}(p)$ is a middle dimensional Lagrangian submanifold of $\mathcal M_H(G,\Sigma_g)$. This fibration is called Hitchin fibration and the base $\mathfrak B$ is the Coulomb branch of 4d class $\mathcal S$ theory by \eqref{fibration3d}. The Hitchin fibration is defined as follows. For each $(A,\varphi)\in\mathcal M_H(G,\Sigma_g)$ we map it to Casimirs of $\varphi$ (which are the invariant $G$ polynomials in $\varphi$):
\be
\pi:(A,\varphi)\mapsto \{\mathrm{Casimirs\, of\,}\varphi\}
\ee
For example when $G=SU(n)$ the Casimirs are $\mathrm{Tr}(\varphi^2),\cdots,\mathrm{Tr}(\varphi^n)$. In general, the base space of Hitchin fibration is:
\be
\mathfrak B=\bigoplus_{i=1}^r H^0(\Sigma_g,K^{d_i})
\ee
where $r$ is the rank of $G$ and $d_i$ are degrees of invariant $G$ polynomials. We can calculate the dimension of $\mathfrak B$ using 
\be
\mathrm{dim}(H^0(\Sigma_g,K^{d_i}))=(g-1)(2d_i-1)
\ee 
Thus the dimension of $\mathfrak B$ is 
\be
\mathrm{dim}(\mathfrak B)= \sum_1^r (g-1)(2d_i-1)=(g-1)\mathrm{dim}(G)
\ee
and the dimension of Hitchin moduli space is $\mathrm{dim}(\mathcal M_H)=2(g-1)\mathrm{dim}(G)$.

From the view of complex geometry $\mathcal M_H$ is a hyperK\"ahler manifold, which means there exists three complex structures $I,J,K$ on $\mathcal M_H$ satisfying:
\be
I^2=J^2=K^2=IJK=-1
\ee 

The Riemannian metric $g$ on $\mathcal M_H$ is K\"ahler with respect to each complex structure. For each complex structure we can define a corresponding K\"ahler form by
\be
\omega_I(v,w)=g(Iv,w)\quad \omega_J(v,w)=g(Jv,w)\quad \omega_K(v,w)=g(Kv,w)
\ee 
where $v,w\in T\mathcal M_H$. The K\"ahler form $\omega_I$ is of type $(1,1)$ with respect to complex structure $I$, and similarly for $\omega_J$ and $\omega_K$. We can further define complex symplectic forms by $\Omega_I=\omega_J+i\omega_K$ and similarly $\Omega_J=\omega_K+i\omega_I $, $\Omega_K=\omega_I+i\omega_J$. Then the symplectic form $\Omega_I$ is of type $(2,0)$ with respect to complex structure $I$. Thus $(\mathcal M_H,I,\Omega_I)$ makes $\mathcal M_H$ a holomorphic symplectic manifold, similarly for $J$ and $K$.

The existence of these complex structures implies that there's a $\mathbb {CP}^1$ family of complex structures on $\mathcal M_H$. Indeed these are:
\be
\mathcal I=aI+bJ+cK\quad a^2+b^2+c^2=1
\ee
We can also define the $\mathbb{CP}^1$ family of complex structures directly by holomorphic coordinates. For each $w\in \mathbb C^*$ we define a complex structure $I_w$ in which $A_{\bar z}-w\varphi_{\bar z}$ and $A_{z}+w^{-1}\varphi_z$ are holomorphic. In this definition the complex structure $I$ corresponds to the point $0\in \mathbb{CP}^1$: $I_0=1$, and similarly $I_{-i}=J$, $I_1=K$.

Note that in complex structure $J$ the holomorphic coordinates are $A_z+i\varphi_z$ and $A_{\bar z}+i\bar\varphi_{\bar z}$, which constitute the $G_{\mathbb C}$-connection $\mathcal A$. Thus in complex structure $J$ the Hitchin moduli space can be identified with the space of flat $G_{\mathbb C}$ connections, and consequently with the character variety: 
\be\label{isomorphictocharacterinJ}
\mathcal M_H(G,\Sigma_g)\simeq\mathcal M_{\mathrm{flat}}(G_{\mathbb C},\Sigma_g)\simeq\mathrm{Hom}(\pi_1(\Sigma_g),G_{\mathbb C})//G_{\mathbb C}
\ee

In summary, straight compactification of class $\mathcal S$ theory on $S^1$ yields a 3d $\mathcal N=4$ sigma model whose target space is Hitchin moduli space. But this statement is not precise as we haven't taken into account the global structures of class $\mathcal S$ theory. If we take into account global structures, the target space is determined as follows \cite{Tachikawa:2013hya}.

A class $\mathcal S$ theory is specified by a simply-laced Lie algebra $\mathfrak g$, a Riemann surface $\Sigma_g$ and a global structure. Let us denote the simply connected Lie group of type $\mathfrak g$ by $\widetilde{G}$ and the center of $\widetilde G$ by $Z(\widetilde G)$. Then the global structure of class $\mathcal S$ theory is a pair of Lagrangian sublattices $(L,L^\bot)$ of $H^1(\Sigma_g,Z(\widetilde G))$ satisfying $L\oplus L^\bot=H^1(\Sigma_g,Z(\widetilde G))$. The Hitchin moduli space $\mathcal M_H(\widetilde G,\Sigma_g)$ is the moduli space of topologically trivial semi-stable $\widetilde G$-Higgs bundles $(E,\varphi)$. For each element $l\in H^1(\Sigma_g,Z(\widetilde G))$ there's an action of $l$ on the $\widetilde G$-bundle $E$. We can identify $l$ with a flat $Z(\widetilde G)$-bundle $\mathcal L(l)$ and tensor $\mathcal L(l)$ to $E$:
\be
l:E\to\mathcal L(l)\otimes E
\ee
We can quotient the moduli space of $\widetilde G$-bundles by the actions of all $l\in H^1(\Sigma_g,Z(\widetilde G))$ to obtain the moduli space of $\widetilde{G}_{\mathrm{adj}}$-bundles, where $\widetilde{G}_{\mathrm{adj}}=\widetilde G/Z(\widetilde G)$. Thus we have:
\be
\mathcal M_H(\widetilde{G},\Sigma_g)/H^1(\Sigma_g,Z(\widetilde G))=\mathcal M_H(\widetilde{G}_{\mathrm{adj}},\Sigma_g)
\ee
For a class $\mathcal S$ theory with global structure $(L,L^\bot)$ we can instead take the quotient of the maximal Lagrangian sublattice $L$:
\be
\mathcal M_H(\widetilde{G},\Sigma_g)/L
\ee
Then the circle compactification of a class $\mathcal S$ theory with global structure $(L,L^\bot)$ is 3d $\mathcal N=4$ sigma model with target space $\mathcal M_H(\widetilde{G},\Sigma_g)/L$.
\subsection{Non-invertible twisted compactification and $(B,B,B)$ branes}\label{section3.2} 
Having reviewed compactification of class $\mathcal S$ theories on $S^1$ with no twists, now we are ready to study \emph{non-invertible twisted compactification} on $S^1$.

\begin{figure}[tbp]
\begin{center}
\[
\hspace*{-0.8 cm}\begin{tikzpicture}[baseline=19,scale=0.8]
\draw[red, thick] (0,-0.2)--(0,3.2);
  \node[left] at (-0.7,1.5) {$p$};
    \node[right] at (+0.7,1.5) {$\mathcal N\cdot p$};
      \node[below] at (0,-0.2) {$\mathcal N$};
\shade[line width=2pt, top color=red,opacity=0.4] 
    (0,0) to [out=90, in=-90]  (0,3)
    to [out=0,in=180] (3,3)
    to [out = -90, in =90] (3,0)
    to [out=190, in =0]  (0,0);
\shade[line width=2pt, color=red,opacity=0.4] 
    (0,0) to [out=90, in=-90]  (0,3)
    to [out=0,in=180] (-3,3)
    to [out = -90, in =90] (-3,0)
    to [out=190, in =0]  (0,0);
\end{tikzpicture}
\hspace{0.35 in}\hspace{0.35 in}
\begin{tikzpicture}[baseline=19,scale=0.8]
\draw[red,  thick] (0,-0.2)--(0,3.2);
\draw[blue, thick] (3,-0.2)--(3,3.2);
  \shade[line width=2pt, color=red,opacity=0.4] 
    (0,0) to [out=90, in=-90]  (0,3)
    to [out=0,in=180] (-3,3)
    to [out = -90, in =90] (-3,0)
    to [out=190, in =0]  (0,0);
          \shade[line width=2pt, top color=red,opacity=0.4] 
    (0,0) to [out=90, in=-90]  (0,3)
    to [out=0,in=180] (3,3)
    to [out = -90, in =90] (3,0)
    to [out=190, in =0]  (0,0);
    \shade[line width=2pt, top color=blue,opacity=0.4] 
    (3,0) to [out=90, in=-90]  (3,3)
    to [out=0,in=180] (6,3)
    to [out = -90, in =90] (6,0)
    to [out=190, in =0]  (3,0);
\node[left] at (-0.7,1.5) {$p$};
    \node at (1.5,1.5) {$\mathcal N_1\cdot p$};
    \node[right] at (3.1,1.5) {$\mathcal N_2\cdot(\mathcal N_1\cdot p)$};
      \node[below] at (0,-0.2) {$\mathcal N_1$};
          \node[below] at (3,-0.2) {$\mathcal N_2$};  
\end{tikzpicture}
\]
\caption{Left: if on one side of $\mathcal N$ the field configuration is $p$, then across the symmetry defect $\mathcal N$ the field configuration must discontinuously change into $\mathcal N\cdot p$. Right: constraints on field configurations imposed by two non-invertible symmetry defects. If $p$ belongs to the moduli space of vacua of non-invertible twisted compactification it must satisfies $p=\mathcal N_1\cdot p=\mathcal N_2\cdot (\mathcal N_1\cdot p).$}
\label{fig:actiononfieldconfig}
\end{center}
\end{figure}
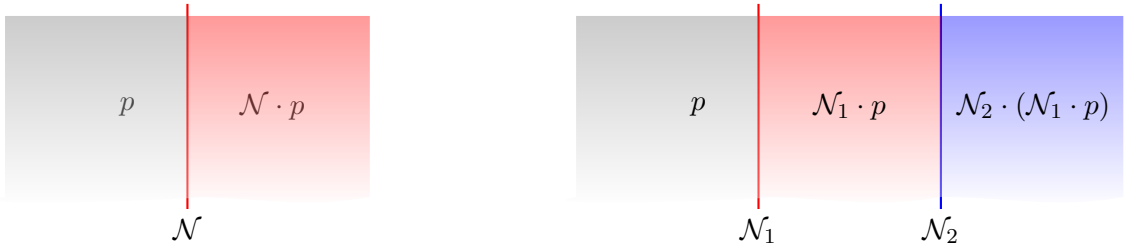

We proceed by first considering the \emph{moduli space of vacua} of resulting 3d theory.  If we add no twists the moduli space of vacua is clearly $\mathcal M=\mathcal M_H(\widetilde{G},\Sigma_g)/L$. A point $p\in\mathcal M$ corresponds to a field configuration of 4d class $\mathcal S$ theory which satisfies equation of motion along $S^1$. If we further add a non-invertible symmetry defect $\mathcal N$, then the defect imposes certain conditions on field configurations of 4d class $\mathcal S$ theory. For example, if on one side of $\mathcal N$ the field configuration is $p$, then across the topological interface $\mathcal N$ the field configuration must become $\mathcal N\cdot p$, as depicted in the left of Fig. \ref{fig:actiononfieldconfig}. Thus if we compactify on $S^1$ in the presence of the non-invertible symmetry defect $\mathcal N$, the moduli space of vacua of resulting 3d theory is the space of field configurations of 4d class $\mathcal S$ theory which not only satisfy equation of motion along $S^1$ but also satisfy the condition imposed by $\mathcal N$.

Note the $\mathcal M=\mathcal M_H(\widetilde{G},\Sigma_g)/L$ is the space of field configurations that satisfy equation of motion along $S^1$. It follows that moduli space of vacua $\mathcal M'$ of non-invertible twisted compactification is the fixed point set of $\mathcal N$ on $\mathcal M$:
\be\label{modulispaceforN}
\mathcal M'=\{p\in\mathcal M|\, \mathcal N\cdot p=p\}
\ee
More generally, if we insert not only one but finitely many non-invertible symmetry defects on $S^1$, they impose similar conditions on field configurations of 4d class $\mathcal S$ theory. If we further compactify on $S^1$, then the moduli space vacua of resulting 3d theory is the subset of $\mathcal M$ consisting of points of $\mathcal M$ that are invariant under each non-invertible symmetry defect. For example, if we insert two non-invertible symmetry defects $\mathcal N_1$, $\mathcal N_2$, the condition on field configurations is illustrated in the right of Fig. \ref{fig:actiononfieldconfig} and the moduli space of vacua of compactification on $S^1$ is:
\be\label{modulispaceforN1N2}
\mathcal M'=\{p\in\mathcal M|\,\mathcal N_1\cdot p=p\quad\mathcal N_2\cdot p=p\}
\ee

Furthermore, as the compactification with no twists yields a sigma model with target space $\mathcal M$, it follows that non-invertible twisted compactification of class $\mathcal S$ theories on $S^1$ is 3d sigma model whose target space is $\mathcal M'$.

Our next goal is to understand $\mathcal M'$ in more detail. To this end we should study action of non-invertible symmetry on the space $\mathcal M=\mathcal M_H(\widetilde{G},\Sigma_g)/L$. Generally a non-invertible symmetry $\mathcal N=F\circ G$ consists of a duality action $F$ and a topological manipulation $G$. The topological manipulation $G$ has no action on the Hitchin moduli space $\mathcal M_H(\widetilde{G},\Sigma_g)$ and only changes the global structure $(L,L^\bot)$. The duality action, which is an element of mapping class group $F\in\mathrm{MCG}(\Sigma_g)$, acts both on $\mathcal M_H(\widetilde{G},\Sigma_g)$ and $(L,L^\bot)$. Furthermore by \eqref{conditionforN} the combined action of $F$ and $G$ on global structure $(L,L^\bot)$ is equal to identity. It follows that $\mathcal N$ only acts on $\mathcal M_H(\widetilde{G},\Sigma_g)$ and has no action on $L$. Thus in the following we neglect the global structure and only study the action of duality $F$ on Hitchin moduli space.

The duality group of class $\mathcal S$ theory is the mapping class group of underlying Riemann surface $\mathrm{MCG}(\Sigma_g)$. For a general Lie group $G$\footnote{In the following we always use $G$ to denote a Lie group but not a topological manipulation.}, how does mapping class group act on Hitchin moduli space $\mathcal M_H({G},\Sigma_g)$? 

By \eqref{isomorphictocharacterinJ}, the Hitchin moduli space in complex structure $J$ is isomorphic to the $G_{\mathbb C}$-character variety:
\be\label{equaltocharacternaive}
\mathcal M_H(G,\Sigma_g)\simeq\mathrm{Hom}(\pi_1(\Sigma_g),G_{\mathbb C})//G_{\mathbb C}
\ee
The mapping class group acts on the fundamental group $\pi_1(\Sigma_g)$ by outer automorphisms. Hence naively by \eqref{equaltocharacternaive} the mapping class group $\mathrm{MCG}(\Sigma_g)$ should have an action on Hitchin moduli space. However note that while the character variety doesn't depend on the complex structure of underlying Riemann surface $\Sigma_g$, a Higgs bundle depends on the complex structure. Thus if a mapping class group action changes the complex structure of $\Sigma_g$, it has no action on Hitchin moduli space. Indeed among the three complex structures $I,J,K$ of $\mathcal M_H(G,\Sigma_g)$, only $J$ doesn't depend on the complex structure of $\Sigma_g$. Similarly only the symlectic form $\Omega_J$ doesn't depend on the complex structure of $\Sigma_g$, while $\Omega_I$ and $\Omega_K$ does.

In order to have an action on Hitchin moduli space we must choose a special complex structure of $\Sigma_g$ for which it is preserved by a subgroup $H\subset \mathrm{MCG}(\Sigma_g)$ of mapping class group. In other words we need to choose the special locus of Teichm\"uller space $\mathrm{Teich}(\Sigma_g)$ at which it is unchanged by a subgroup $H$ of mapping class group. Indeed this requirement coincides with the condition \eqref{conditionforN} for non-invertible symmetry: the Teichm\"uller space is the conformal manifold of class $\mathcal S$ theory; in order to have non-invertible symmetry we need to choose a self-dual point of conformal manifold which is preserved by some duality action.

Having chosen a suitable locus of Teichm\"uller space, the stablizer subgroup of mapping class group $H$ should have an action on Hitchin moduli space $\mathcal M_H(G,\Sigma_g)$. For a particular element $\varphi$ of $H$ we may construct a non-invertible symmetry $\mathcal N$. Then by \eqref{modulispaceforN} the target space of non-invertible twisted compactification is:
\be
\mathcal M'=\{x\in\mathcal M_H(G,\Sigma_g)|\,\varphi(x)=x\}\footnote{More precisely we should quotient $\mathcal M'$ by $L$. But as we've said above, now we neglect global structures.}
\ee 
More generally, if we insert more than one non-invertible symmetry defects, by \eqref{modulispaceforN1N2} we obtain a submanifold $\mathcal M'$ of Hitchin moduli space, labeled by a subgroup of mapping class group $H$:
\be
\mathcal M'=\{x\in\mathcal M_H(G,\Sigma_g)|\,\varphi(x)=x\quad\forall \varphi\in H\}
\ee

What is the geometry of the submanifold $\mathcal M'$? As defined in \cite{Kapustin:2006pk}, a submanifold of Hitchin moduli space is called a \emph{$B$-brane} if it is a complex submanifold in complex structure $I$, and it's called a \emph{$A$-brane} if it is a Lagrangian submanifold in the symplectic form $\omega_I$, similarly for $J$ and $K$. For each of the three complex structures $I,J,K$ we may ask whether a submanifold is a $B$-brane or an $A$ brane and we may get, e.g. $(B,B,B)$ branes and $(A,B,A)$ branes.

In particular a $(B,B,B)$ brane is a complex submanifold in all complex structures, thus a hyperK\"ahler submanifold of Hitchin moduli space. If $\mathcal M'$ is a $(B,B,B)$ brane, then the 3d sigma model with target space $\mathcal M'$ has $\mathcal N=4$ supersymmetry. How can we prove that $\mathcal M'$ is a $(B,B,B)$ brane?

In general, a class of $(B,B,B)$ branes of Hitchin moduli space can be constructed via finite groups acting on the Riemann surface $\Sigma_g$ \cite{Heller:2016drq,Biswas:2020znv,Schaposnik:2019xas}. Consider a finite group $\Gamma$ acting on the Riemann surface $\Sigma_g$ by holomorphic automorphisms. For an element $\Gamma\ni\tau: \Sigma_g\to\Sigma_g$ it induces an action on the Hitchin moduli space $\mathcal M_H(G,\Sigma_g)$ by pullback:
\be
\tau^*: (E,\varphi)\to(\tau^*E,\tau^*\varphi)
\ee
The theorem of \cite{Heller:2016drq} is that:
\begin{thm}
For a Riemann surface $\Sigma_g$ of genus $g\geq2$ and a finite group $\Gamma$ acting on $\Sigma_g$ by holomorphic automorphisms, the fixed point set of $\Gamma$ on Hitchin moduli space $\mathcal M_H(G,\Sigma_g)$ is a $(B,B,B)$ brane.
\end{thm} 
 
How to apply this theorem to our problem? Note that in our problem the submanifold $\mathcal M'$ is the fixed point set of a subgroup $H$ of mapping class group that has fixed points on Teichm\"uller space $\mathrm{Teich}(\Sigma_g)$. By Nielsen realization theorem \cite{Kerckhoff1980TheNR,NAKAMURA20183585}, for $g\geq 2$ the following statements are equivalent:
\begin{enumerate}
\item $H$ is a finite subgroup of mapping class group $\mathrm{MCG}(\Sigma_g)$.
\item $H$ is a subgroup of mapping class group $\mathrm{MCG}(\Sigma_g)$ that has fixed points on Teichm\"uller space $\mathrm{Teich}(\Sigma_g)$.
\item $H$ is a subgroup of mapping class group $\mathrm{MCG}(\Sigma_g)$ that can be realized as automorphisms on the surface $\Sigma_g$.
\end{enumerate}
Thus combining Nielsen realization theorem with the above theorem for $(B,B,B)$ branes, we have proved that the target space $\mathcal M'$ of non-invertible twisted compactification is a $(B,B,B)$ brane, i.e. a hyerK\"ahler submanifold\footnote{I thank Nigel J. Hitchin for pointing out this to me.}.

In summary, non-invertible twisted compactification of class $\mathcal S$ theories on $S^1$ yields 3d $\mathcal N=4$ sigma model whose target space is a $(B,B,B)$ brane of Hitchin moduli space. The $(B,B,B)$ brane is the fixed point set of a finite subgroup of mapping class group on Hitchin moduli space.
\section{$(B,B,B)$ brane as an affine variety}\label{section4}
\subsection{Loop coordinates}\label{section4.1}
In the previous section we have seen non-invertible twisted compactification of class $\mathcal S$ theory on $S^1$ is sigma model with target space a $(B,B,B)$ brane $\mathcal M'$. The aim of this section is to describe the $(B,B,B)$ brane as an \emph{affine variety}. This description of $(B,B,B)$ branes, to the author's knowledge, has never appeared in mathematical literature. More concretely we want to calculate type $A_1$, genus $2$ case, for which the symmetry generators are collected in Table \ref{tab:fixedpoints} and Appendix \ref{appendixA}. In this subsection we describe the Hitchin moduli space $\mathcal M_H(G,\Sigma_g)$ in terms of \emph{loop coordinates}, in which the Hitchin moduli space is expressed as the zero locus of some polynomials on them.

Recall that the Hitchin moduli space $\mathcal M_H(G,\Sigma_g)$ in complex structure $J$ and symplectic form $\Omega_J$ is isomorphic to the $G_{\mathbb C}$-character variety. For type $A_{n+1}$ the Lie group is $G=SU(n)$ and the complexification is $G_{\mathbb C}=SL(n,\mathbb C)$. The fundamental group of the Riemann surface of genus $g$ is:
\be
\pi_1(\Sigma_g)=\{X_1,Y_1\cdots X_g,Y_g|\,\prod_{i=1}^g[X_i,Y_i]=1\}
\ee
where $[X_i,Y_i]=X_i Y_iX_i^{-1}Y_i^{-1}$ is the commutator of between two group elements. An element of the character variety $\rho\in\mathrm{Hom}(\pi_1(\Sigma_g),SL(n,\mathbb C))//SL(n,\mathbb C)$ assigns a matrix $\rho(X_i)\in SL(n,\mathbb C)$ to each generator $X_i$ of the fundamental group. For simplicity we will denote the matrix $\rho(X_i)$ simply by the generator $X_i$. Then the character variety is the space of $2g$ $SL(n,\mathbb C)$ matrices $\{X_i,Y_i\}$ satisfying:
\be
\prod_{i=1}^gX_i Y_iX_i^{-1}Y_i^{-1}=1
\ee
modulo a simultaneous $SL(n,\mathbb C)$ conjugation. Actually the character variety is an affine variety, which means it can be defined as the zero locus of some polynomials. Indeed we can parametrize the character variety by all matrix elements $X^{(i,j)}_k,Y^{(i,j)}_k$ of $X_k$ and $Y_k$. For each matrix element of $\prod_{i=1}^gX_i Y_iX_i^{-1}Y_i^{-1}$ we get an polynomial relation on the coordinates $X^{(i,j)}_k,Y^{(i,j)}_k$. Together with $2g$ polynomials $\mathrm{det}(X_k)-1$ and $\mathrm{det}(Y_k)-1$, the zero locus of these polynomials defines the representation variety $\mathrm{Hom}(\pi_1(\Sigma_g),SL(n,\mathbb C))$. For each affine variety there's a corresponding ring called \emph{coordinate ring}. It's defined as follows. If the polynomials defining the affine variety are on $m$ variebles, they are elements in $\mathbb C[x_1,\cdots,x_m]$. Denote the ideal generated by these polynomials by $\mathcal I\subset \mathbb C[x_1,\cdots,x_m]$. Then the coordinate ring $\mathcal A$ is simply defined as
\be
\mathcal A=\mathbb C[x_1,\cdots,x_m]/\mathcal I
\ee
The affine variety can be recovered from the coordinate ring. Indeed the variety is just the spectrum of coordinate ring $\mathrm{Spec}(\mathcal A)$. When it comes to character variety, its coordinate ring is the $SL(n,\mathbb C)$ invariant subring of the coordinate ring of representation variety. In the following we will denote by $\mathcal A$ the coordinate ring of character variety. 

As the ring of holomorphic functions on character variety, $\mathcal A$ is endowed with a Poisson algebra structure. Indeed in terms of the symplectic form $\Omega_J$ we can calculate the Poisson bracket between two elements of $\mathcal A$. For type $A_1$ which corresponds to $SL(2,\mathbb C)$ character variety, the quantization of this Poisson algebra is known to be the skein algebra $Sk_q(\Sigma_g)$ of the surface \cite{Turaev1991,kabiraj2022centerpoissonskeinalgebras}. There is a general theorem that the mapping class group $\mathrm{MCG}(\Sigma_g)$ always acts on the coordinate ring $\mathcal A$ of character variety by \emph{automorphisms} of its Poisson algebra. Thus the action of mapping class group $\mathrm{MCG}(\Sigma_g)$ on character variety is holomorphic and preserves the holomorphic symplectic form $\Omega_J$. Hence the fixed point set of a subgroup $H$ of mapping class group is a holomorphic symplectic submanifold, i.e. a hyperK\"ahler submanifold. The coordinate ring of the fixed point set is just the invariant subring of $\mathcal A$ for the subgroup $H$ of mapping class group $\mathrm{MCG}(\Sigma_g)$. Thus we get a description of the $(B,B,B)$ brane as an affine variety.

As a ring, it's known that $\mathcal A$ is  generated by the traces $\{\mathrm{tr}(X)|\,x\in\pi_1(\Sigma_g)\}$. We will call the coordinates $\{\mathrm{tr}(X)\in\mathbb C\,|\,x\in\pi_1(\Sigma_g)\}$ \emph{loop coordinates}, named in \cite{Dimofte:2011jd}. The loop coordinates make the computation of the action of mapping class group $\mathrm{MCG}(\Sigma_g)$ on character variety very easy. Furthermore, the traces $\{\mathrm{tr}(X)|\,x\in\pi_1(\Sigma_g)\}$ are not all independent. For $SL(2,\mathbb C)$ character variety it's known that the traces $\{\mathrm{tr}(X)|\,x\in\pi_1(\Sigma_g)\}$ are generated by $2^{2g}-1$ traces \cite{Magnus1980}. These traces are defined as follows. The group $\pi_1(\Sigma_g)$ has $2g$ generators $\{X_i,Y_i\quad i=1,\cdots g\}$. The $2^{2g}-1$ traces can be defined as:
\be
\mathrm{tr}(X_{\mu_1}\cdots X_{\mu_{k_1}}Y_{\nu_1}\cdots Y_{\nu_{k_2}})\quad \mu_1\leq \mu_2\leq\cdots\leq\mu_{k_1}\quad \mu_1\leq \nu_2\leq\cdots\leq\nu_{k_1}
\ee
The number of combinations of this form is just 
\be
\sum_{n=1}^{2g} C_{2g}^n-1=2^{2g}-1
\ee
Let $W(X_i,Y_i)$ be a word on the generators $\{X_i,Y_i\}$. Then it's known that $\mathrm{tr}(W)$ can be represented as a polynomial on the $2^{2g}-1$ traces with integral cofficients \cite{VorlesungenD,Horowitz1972CharactersOF}. 

Having finished the general arguments to calculate the $(B,B,B)$ branes, let us compute some concrete examples. We compute the $SL(2,\mathbb C)$ character variety of one-punctured torus \cite{tan2005sl2ccharactervarietyoneholed,Nekrasov:2011bc,Dimofte:2011jd}. For the one-punctured torus its fundamental group is:
\be
\pi_1(\Sigma_{1,1})=\{X,Y,Z|\, ZXYX^{-1}Y^{-1}=1\}
\ee
where $Z$ is the cycle around the puncture. For punctured Riemann surface its character variety is defined by specifying the conjugacy class for holonomies around puncture. For example for $\Sigma_{1,1}$ we should fix the conjugacy class of the matrix $Z$. For $G_{\mathbb C}=SL(2,\mathbb C)$ the conjugacy class of an element $Z$ is labeled by the trace $\mathrm{tr}(Z)$. Thus we need to fix $\mathrm{tr}(Z)=m$. The character variety is just the space of three $SL(2,\mathbb C)$ matrices $X_1,Y_1,Z$ which satisfies
\be
ZXYX^{-1}Y^{-1}=1
\ee
up to a simultaneous $SL(2,\mathbb C)$ conjugation. Combining it with requiring $\mathrm{tr}(Z)=m$, we see that the character variety is the space of two $SL(2,\mathbb C)$ matrices $X,Y$ satisfying 
\be
\mathrm{tr}(XYX^{-1}Y^{-1})=m
\ee
where we have used the fact that for $SL(2,\mathbb C)$ matrices, $\mathrm{tr}(Z)=\mathrm{tr}(Z^{-1})$. 

Now we define loop coordinates for the two generators $X,Y$:
\be
x_1:=\mathrm{tr}(X)\quad x_2:=\mathrm{tr}(Y)\quad x_3:=\mathrm{tr}(XY)
\ee 
Indeed they are the $2^{2}-1=3$ traces that generate the traces of full fundamental group. For example can transform $\mathrm{tr}(XYX^{-1}Y^{-1})$ into a polynomial on the three loop coordinates. By repeated using the crucial identity for $SL(2,\mathbb Z)$ matrices:
\be
\mathrm{tr}(A)\mathrm{tr}(B)=\mathrm{tr}(AB)+\mathrm{tr}(AB^{-1})
\ee
we have 
\be
\begin{aligned}
\mathrm{tr}(XYX^{-1}Y^{-1})&=\mathrm{tr}(XY)\mathrm{tr}(X^{-1}Y^{-1})-\mathrm{tr}(XY^2X)=x_3^2-\mathrm{tr}(X^2Y^2)\\
\mathrm{tr}(X^2Y^2)&=\mathrm{tr}(X^2Y)\mathrm{tr}(Y)-\mathrm{tr}(X^2)=\mathrm{tr}(X^2Y)x_2-\mathrm{tr}(X^2)\\
\mathrm{tr}(X^2Y)&=\mathrm{tr}(X)\mathrm{tr}(XY)-\mathrm{tr}(Y)=x_1x_3-x_2\\
\mathrm{tr}(X^2)&=\mathrm{tr}(X)\mathrm{tr}(X)-\mathrm{tr}(1)=x_1^2-2
\end{aligned}
\ee
Finally we work out that the character variety $\chi_{1,1}$ for $\Sigma_{1,1}$ is a cubic surface in $\mathbb C^3$:
\be
\chi_{1,1}=\{(x_1,x_2,x_3)\in\mathbb C^3\,|\, W_{1,1}=x_1^2 +x_2^2+x_3^2+x_1x_2x_3-2-m=0\}
\ee

Besides we can easily work out the action of mapping class group on $\chi$. The mapping class group for one-punctured torus is $SL(2,\mathbb Z)$. Let's take the $S$ transformation for an example: $S=\left(\begin{matrix}0&-1\\1&0\\\end{matrix}\right)\in SL(2,\mathbb Z)$. It acts on generators of fundamental group by 
\be
\begin{aligned}
X&\mapsto Y^{-1}\\
Y&\mapsto X
\end{aligned}
\ee
It follows that the action on loop coordinates is:
\be\label{actionofSonloop}
\begin{aligned}
x_1&=\mathrm{tr}(X)\mapsto \mathrm{tr}(Y^{-1})=x_2\\
x_2&=\mathrm{tr}(Y)\mapsto \mathrm{tr}(X)=x_1\\
x_3&=\mathrm{tr}(XY)\mapsto \mathrm{tr}(Y^{-1}X)=\mathrm{tr}(Y^{-1})\mathrm{tr}(X)-\mathrm{tr}(YX)=x_1x_2-x_3\\
\end{aligned}
\ee
We can easily verify that this action preserves the polynomial $W_{1,1}$. And the fixed point set of $S$ transformation is simply:
\be
\begin{aligned}
&x_1^2 +x_2^2+x_3^2+x_1x_2x_3-2-m=0\\
&x_1=x_2\\
&2x_3=x_1x_2\\
\end{aligned}
\ee
It's easily calculated that the solution to above equation is:
\be
\{x_1\in\mathbb C \,|\,3x^4_1+8x^2_1-8-4m=0\}
\ee
This manifold has complex dimension 0, thus is a zerobrane of Hitchin moduli space. Indeed zerobranes are $(B,B,B)$ branes.

Besides, the loop coordinates are functions on the character variety. We can calculate the Poission bracket between them in terms of the holomorphic symplectic form $\Omega_J$. And the result is \cite{Nekrasov:2011bc,Dimofte:2011jd}:
\be
\begin{aligned}
\{x_1,x_2\}&=x_3-\frac{1}{2}x_1x_2\\
\{x_1,x_3\}&=-x_2+\frac{1}{2}x_1x_3\\
\{x_2,x_3\}&=x_1-\frac{1}{2}x_2x_3\\
\end{aligned}
\ee
It's straightforward to verify that the $S$ transformation \eqref{actionofSonloop} preserves the Poission algebra. Thus the action of mapping class group preserves the holomorphic symplectic form $\Omega_J$ and the fixed point set of this action is a holomorphic symplectic submanifold, i.e. a hyperK\"ahler submanifold. 
\subsection{Genus 2 case}\label{section4.2}
Having reviewed loop coordinates of character variety. Now we calculate the $(B,B,B)$ branes corresponding to non-invertible twisted compactification of genus 2 class $\mathcal S$ theory of type $A_1$. In this case the symmetry generators are listed in Table \ref{tab:fixedpoints} and Appendix \ref{appendixA}.

Let us first describe the character variety $\chi_{2,0}$ of $\Sigma_{2,0}$ in terms of loop coordinates. This is calculated in \cite{arthamonov2024classicallimitgenusdaha}, see also \cite{Arthamonov_2019,Cooke_2021}.

The genus 2 fundamental group $\pi_1(\Sigma_{2,0})$ has $4$ generators $X_1,Y_1,X_2,Y_2$ and they satisfy the relation:
\be
\pi_1(\Sigma_{2,0})=\{X_1,Y_1,X_2,Y_2\,|\,\prod_{i=1}^2 X_i Y_iX_i^{-1}Y_i^{-1}=1\}
\ee

We define $15$ loop coordinates:
\be
(z_1,z_2,z_3,z_4,z_5,z_6,z_{12},z_{23},z_{34},z_{45},z_{56},z_{61},z_{123},z_{234},z_{345})\in\mathbb C^{15}
\ee
In terms of the generators $\{X_1,Y_1,X_2,Y_2\}$ they are defined as:
\be
\begin{aligned}
z_1=&\mathrm{tr}(Y_1),
&z_{12}=&\mathrm{tr}(Y_1X_1^{-1}),
&z_{123}=&\mathrm{tr}(Y_1X_1^{-1}Y_1^{-1}Y_2^{-1}),\\
z_2=&\mathrm{tr}(X_1),
&z_{23}=&\mathrm{tr}(X_1^{-1}Y_1^{-1}Y_2^{-1}),
&z_{234}=&\mathrm{tr}(X_1Y_1Y_2^2X_2^{-1}Y_2^{-1})\\
z_3=&\mathrm{tr}(Y_1Y_2),
&z_{34}=&\mathrm{tr}(Y_2^{-1}Y_1^{-1}X_2),
&z_{345}=&\mathrm{tr}(X_2Y_1^{-1})\\
z_4=&\mathrm{tr}(X_2),
&z_{45}=&\mathrm{tr}(Y_2X_2),\\
z_5=&\mathrm{tr}(Y_2),
&z_{56}=&\mathrm{tr}(X_1Y_2^2X_2^{-1}Y_2^{-1}),\\
z_6=&\mathrm{tr}(X_1Y_2X_2^{-1}Y_2^{-1}),
&z_{61}=&\mathrm{tr}(X_1Y_1Y_2X_2^{-1}Y_2^{-1}),
\end{aligned}
\ee

The character variety $\chi_{2,0}\subset\mathbb C^{15}$ can be defined as the zero locus of some polynomials on the 15 variebles. As calculated in \cite{arthamonov2024classicallimitgenusdaha}, there're 19 polynomial relations on the $15$ loop coordinates. These are: 
\begin{align}
&z_{i+2}z_{i+4}+z_{i+3}z_{i+2,i+3,i+4}-z_{i+2,i+3} z_{i+3,i+4}-2z_i=0 \nonumber\\
&z_{i+3}z_{i+5,i} +z_{i+4}z_{i+1,i+2}-z_{i+3,i+4} z_{i+1,i+2,i+3}+2( z_{i,i+1}-z_iz_{i+1})=0 \nonumber\\
&z_{i+1,i+2}z_{i+4,i+5} -z_{i,i+1,i+2} z_{i+1,i+2,i+3}+z_iz_{i+3} -2(z_{i+2,i+3,i+4}- z_{i+1}z_{i+5,i}-z_{i+5}z_{i,i+1}\nonumber\\
&\quad\quad\quad\quad\quad\quad\quad\quad\quad  +z_{i+5}z_{i,i+1}+z_{i+5}z_{i,i+1}+z_iz_{i+1}z_{i+5})=0\label{relationforgenus2character}\\
&z_{123}z_{234}z_{345} -(z_1z_4z_{345} +z_2z_5z_{123} +z_3z_6z_{234})+(z_1z_6z_{61} +z_2z_1z_{12} +z_3z_2z_{23} +z_4z_3z_{34} +\nonumber\\
&\quad z_5z_4z_{45} +z_6z_5z_{56})-(z_1z_3z_5 +z_2z_4z_6)-(z_{12}^2+z_{23}^2+z_{34}^2+z_{45}^2+z_{56}^2+z_{61}^2)+8=0\nonumber
\end{align}
where $1\leq i\leq 6$. The Hitchin moduli space $\mathcal M_H(SU(2),\Sigma_2)=\chi_{2,0}\subset \mathbb C^{15}$ is the zero locus of these polynomials. \cite{arthamonov2024classicallimitgenusdaha} proves that the zero locus has complex dimension $6$. Indeed this coincides with the dimension of Hitchin moduli space:
\be
\mathrm{dim}(\mathcal M_H(SU(2),\Sigma_2))=2(g-1)\,\mathrm{dim}(G)=2\,\mathrm{dim}(SU(2))=6
\ee 

For any word $W(X_1,Y_1,X_2,Y_2)$ on the $4$ generators, its trace $\mathrm{tr}(W)$ is a polynomial on the 15 loop coordinates. For example we can explicitly calculate: 

\begin{align}
\mathrm{tr}(X_1)=& z_2,&
\mathrm{tr}(Y_2)=& z_5,&
\mathrm{tr}(X_1Y_1)=& z_1 z_2-z_{12},\nonumber\\
\mathrm{tr}(Y_1)=& z_1,&
\mathrm{tr}(Y_1Y_2)=& z_3,&
\mathrm{tr}(Y_1X_2)=& z_1 z_4-z_{345},\nonumber\\
\mathrm{tr}(X_2)=& z_4,&\mathrm{tr}(X_2Y_2)=& z_{45},&
\mathrm{tr}(Y_1X_2Y_2)=& z_{34}+z_1 z_{45}-z_5 z_{345},\nonumber
\end{align}
\begin{align}
\mathrm{tr}(X_1X_2)=& -z_1 z_{61}+z_1 z_2 z_{345}-z_{12} z_{345}+z_6,\label{eq:PhiActionOnGenerators}\\
\mathrm{tr}(X_1Y_2)=& z_3 z_{12}+z_1 z_{23}-z_{123}-z_1 z_2 z_3+z_2 z_5,\nonumber\\
\mathrm{tr}(X_1Y_1X_2)=&- z_1^2 z_{61}+z_2 z_1^2 z_{345}-z_1 z_{12} z_{345}+z_{61}-z_2 z_{345}+z_6 z_1,\nonumber\\
\mathrm{tr}(X_1Y_1Y_2)=& -z_5 z_{12}-z_4 z_5 z_{61}-z_{23}+z_{45} z_{61}+ z_4 z_{234}+z_1 z_2 z_5,\nonumber\\
\mathrm{tr}(X_1X_2Y_2)=& -z_1 z_2 z_{34}+z_{12} z_{34}-z_{56}-z_1 z_5 z_{61}+z_1 z_{234}+z_1 z_2 z_5 z_{345}-z_5 z_{12} z_{345}+z_5 z_6.\nonumber
\end{align}

In order to study $(B,B,B)$ branes associated to finite subgroups of mapping class group, we should study the action of mapping class group $\mathrm{MCG}(\Sigma_2)$ on the character variety $\chi_{2,0}$. The mapping class group of genus 2 $\mathrm{MCG}(\Sigma_2)$ is generated by $5$ Dehn twists $d_1,d_2,d_3,d_4,d_5$ satisfying the relations \cite{Birman1971OnTM}:
\be\label{relation1genus2}
\begin{gathered}
d_id_j=d_jd_i\quad |i-j|> 1 \\
d_id_{i+1}d_i=d_{i+1}d_id_{i+1}\quad 1\leq i\leq 4\\
\end{gathered}
\ee
and
\be
I^6=H^2=1\quad \mathrm{for}\quad I=d_1d_2d_3d_4d_5\quad H=d_1d_2d_3d_4d_5d_5d_4d_3d_2d_1
\ee

Note that by \eqref{relation1genus2} we have
\be
d_i=I^{i-1}d_1 I^{1-i}\quad 1\leq i\leq 5
\ee
It follows that the mapping class group is generated two elements $\{d_1,I\}$. The two elements act on the fundamental group by:
\begin{align}
d_1:\quad\left\{\begin{array}{ccl}
X_1&\mapsto& X_1Y_1\\
Y_1&\mapsto& Y_1\\
X_2&\mapsto& X_2\\
Y_2&\mapsto& Y_2
\end{array}\right.
\qquad\qquad\qquad
I:\quad\left\{\begin{array}{ccl}
X_1&\mapsto& Y_2Y_1\\
Y_1&\mapsto& X_1^{-1}\\
X_2&\mapsto& (X_1^{-1}Y_2X_2)Y_2(X_1^{-1}Y_2X_2)^{-1}\\
Y_2&\mapsto& Y_2(X_1^{-1}Y_2X_2)^{-1}
\end{array}\right.
\end{align}

From this we can work out the action of $d_1$ and $I$ on loop coordinates. For example the action of $I$ on $z_2$ is:
\be
z_2=\mathrm{tr}(X_1)\mapsto \mathrm{tr}(Y_2Y_1)=z_3
\ee
More generally we can verify that $I$ acts on the loop coordinates by permutation:
\begin{align}
I:\quad\left\{\begin{array}{ccl}
z_1\mapsto z_2\mapsto z_3\mapsto z_4\mapsto z_5\mapsto z_6\mapsto z_1\\
z_{12}\mapsto z_{23}\mapsto z_{34}\mapsto z_{45}\mapsto z_{56}\mapsto z_{61}\mapsto z_{12}\\
z_{123}\mapsto z_{234}\mapsto z_{345}\mapsto z_{123}
\end{array}\right.
\end{align}
 
Similarly we can calculate the action of $d_1$ and the result is \cite{arthamonov2024classicallimitgenusdaha}:
\begin{align}
d_1:\quad\left\{\begin{array}{cll}
z_2 &\mapsto &  z_1z_2- z_{12} \\[2pt]
z_6 &\mapsto& z_{61} \\[2pt]
z_{12} &\mapsto& z_2 \\[2pt]
z_{23} &\mapsto& z_1z_{23}- z_{123} \\[2pt]
z_{56} &\mapsto& z_{234} \\[2pt]
z_{61} &\mapsto&  z_1z_{61}-z_6  \\[2pt]
z_{123} &\mapsto& z_{23} \\[2pt]
z_{234} &\mapsto& z_1z_{234}- z_{56}
\end{array}\right.
\end{align} 
and $d_1$ acts trivially on other loop coordinates.

 From the action of $I$ and $d_1$ we can obtain the action of whole mapping class group on character variety. Among the elements of mapping class group, only the ones having finite order have an action on Hitchin moduli space by Nielsen realization theorem. If we fix a finite subgroup $H$ of mapping class group we can study its fixed point set on Hitchin moduli space, which is the $(B,B,B)$ brane of interest. For example, the element $I$ has order $6$. Thus we can study the fixed point set of the $\mathbb Z_6$ subgroup generated by $I$ and it is:

\begin{align}
\left\{\begin{array}{ccl}
z_1= z_2= z_3= z_4= z_5= z_6\\
z_{12}= z_{23}= z_{34}= z_{45}= z_{56}= z_{61}\\
z_{123}= z_{234}= z_{345}
\end{array}\right.
\end{align}
Together with \eqref{relationforgenus2character} these polynomials defines an affine variety, which is the $(B,B,B)$ brane.

How can we prove that these affine varieties are indeed hyperK\"ahler manifolds? The holomorphic symplectic form $\Omega_J$ on the character variety $\chi_{2,0}$ induces a Poisson algebra for the loop coordinates. The mapping class group acts as automorphisms of this Poisson algebra \cite{arthamonov2024classicallimitgenusdaha}. Thus the action is holomorphic and preserves the holomorphic symplectic form. Consequently the fixed point set is a holomorphic symplectic submanifold. Actually the result of  \cite{arthamonov2024classicallimitgenusdaha} is stronger: it considers a two-parameter deformation $\mathcal A_{q,t}$ of Poisson algebra $\mathcal A$ (called \emph{genus 2 double affine Hecke algebra}) and proves that mapping class group acts by automorphisms of $\mathcal A_{q,t}$. Thus the $(B,B,B)$ branes considered in this paper have very rich algebraic structures. It would be interesting to study how the algebraic structure is related to physics.
\section{Conclusion and outlook}

In this paper we have studied non-invertble twisted compactification of class $\mathcal S$ theories on $S^1$. We have shown that the resulting 3d theory is 3d $\mathcal N=4$ sigma model whose target space $\mathcal M'$ is a $(B,B,B)$ brane of Hitchin moduli space. The $(B,B,B)$ branes are fixed point sets of finite subgroups of mapping class group of underlying Riemann surface. For type $A_1$ genus $2$ class $\mathcal S$ theory with no punctures, we describe the $(B,B,B)$ branes as the zero locus of some polynomials, i.e. an affine variety. Our calculation exhibits that the $(B,B,B)$ branes have very rich algebraic structures.  

There are many directions to further study. 
\begin{enumerate}
\item In this paper we have only considered the Coulomb branch of the resulting 3d theory. The Higgs branch of it needs to be further studied. Generally for a 3d $\mathcal N=4$ SCFT $\mathcal T$ there is a mirror theory $\mathcal T^\vee$ whose Coulomb branch is the Higgs branch of $\mathcal T$ and whose Higgs branch is the Coulomb branch of $\mathcal T$. This operation is called \emph{3d mirror symmetry} \cite{Intriligator:1996ex}. For circle compactification of class $\mathcal S$ theory with no twists, its 3d mirror dual is star-shaped quiver gauge theory \cite{Benini:2010uu}.  It would be interesting to study what is the 3d mirror theory for non-invertible twisted compactification of class $\mathcal S$ theory. 
\item For each 3d $\mathcal N=4$ sigma model whose target space is $\mathcal M$, we can perform a topological twist and obtain a 3d topological field theory. Since $\mathcal M$ is always a hyperK\"ahler manifold, this operation defines a 3d TFT for each hyperK\"ahler manifold. The corresponding 3d TFT is called \emph{Rozansky-Witten theory} \cite{Rozansky:1996bq,Gukov:2020lqm,Gukov:2024adb}. It would be interesting to study Rozansky-Witten theory for the $(B,B,B)$ branes proposed in this paper.
\item While non-invertible symmetry has become popular among physics community, non-invertible twisted compactification has rarely been studied. It would be interesting to study non-invertible twisted compactification for more general 4d $\mathcal N=2$ SCFTs and discover whether they share similar structures in common.
\item While the $(B,B,B)$ branes proposed in this paper have both rich geometric and algebraic structures, it has rarely been studied in mathematical literature. It would be interesting to understand their mathematical structures in more detail. 
\end{enumerate}
\subsection*{Acknowledgements}

I thank Semeon Arthamonov, Nigel J. Hitchin and Yiwen Pan for useful discussions.
\clearpage
\appendix
\section{Generators of genus 2}\label{appendixA}
In this appendix we give explicit matrix representation of symmetry generators of genus 2 appearing in Table \ref{tab:fixedpoints}.
\be
\begin{gathered}
\phi=\left(\begin{matrix}0&-1&-1&-1\\0&0&-1&0\\0&0&0&-1\\1&0&0&1\\\end{matrix}\right)
,\quad M_1=\left(\begin{matrix}0&0&0&1\\0&0&1&1\\1&-1&0&0\\-1&0&0&0\\\end{matrix}\right),\\
M_2=\left(\begin{matrix}0&0&1&1\\0&0&1&0\\0&-1&0&0\\-1&1&0&0\\\end{matrix}\right)
,\quad M_3=\left(\begin{matrix}0&0&0&1\\0&0&-1&0\\0&-1&0&0\\1&0&0&0\\\end{matrix}\right),\\
M_4=\left(\begin{matrix}0&0&1&0\\0&0&0&1\\-1&0&-1&0\\0&-1&0&0\\\end{matrix}\right)
,\quad M_5=\left(\begin{matrix}0&0&1&0\\0&-1&0&0\\-1&0&0&0\\0&0&0&-1\\\end{matrix}\right),\\
M_6=\left(\begin{matrix}0&0&-1&0\\0&0&0&1\\1&0&0&0\\0&-1&0&0\\\end{matrix}\right)
,\quad M_7=\left(\begin{matrix}0&1&0&0\\1&0&0&0\\0&0&0&1\\0&0&1&0\\\end{matrix}\right),\\
M_8=\left(\begin{matrix}-1&1&1&0\\1&0&0&1\\-1&0&0&0\\1&-1&0&1\\\end{matrix}\right)
,\quad M_9=\left(\begin{matrix}0&0&0&-1\\1&0&1&0\\0&1&0&1\\-1&0&0&0\\\end{matrix}\right),\\
M_{10}=\left(\begin{matrix}0&0&1&0\\0&0&0&-1\\-1&0&-1&0\\0&1&0&1\\\end{matrix}\right)
,\quad N_1=\left(\begin{matrix}0&0&1&0\\0&1&0&0\\-1&0&0&0\\0&0&0&1\\\end{matrix}\right),\\
N_2=\left(\begin{matrix}0&1&-1&0\\-1&0&0&1\\0&0&0&1\\0&0&-1&0\\\end{matrix}\right)
,\quad N_3=\left(\begin{matrix}0&-1&0&0\\1&0&0&0\\0&0&0&-1\\0&0&1&0\\\end{matrix}\right),\\
N_4=\left(\begin{matrix}1&0&1&0\\0&1&0&0\\-1&0&0&0\\0&0&0&1\\\end{matrix}\right)
,\quad N_5=\left(\begin{matrix}-1&1&0&0\\0&1&0&0\\0&0&-1&0\\0&0&1&1\\\end{matrix}\right),\\
N_6=\left(\begin{matrix}1&0&0&0\\1&-1&0&0\\0&0&1&1\\0&0&0&-1\\\end{matrix}\right)
,\quad C=\left(\begin{matrix}-1&0&0&0\\0&-1&0&0\\0&0&-1&0\\0&0&0&-1\\\end{matrix}\right),\quad
P=\left(\begin{matrix}1&0&0&0\\0&-1&0&0\\0&0&1&0\\0&0&0&-1\\\end{matrix}\right).\\
\end{gathered}
\ee
\bibliographystyle{JHEP}
\bibliography{nontwistclassS}
\end{document}